\documentstyle[prd,aps]{revtex}
\begin{document}
\input psfig


\title{
\quad
  \large \hfill CLNS 96/1454  \\
  \large \hfill CLEO 96-23
\quad \\
\vspace{1.2cm}
\Large \bf The Inclusive Decays 
$B\rightarrow D X $ and $B\rightarrow D^* X$  \\
\vspace{1.2cm}
\large  CLEO collaboration  \\
\date{\today}
}  
\maketitle

\begin{abstract}
\vspace{1cm}  
\begin{center}
\large Abstract 
\end{center}
\vspace{1cm}  
We report new measurements of the differential and total branching ratios for
inclusive $B$ decay to $D^0$, $D^+$ and $D^{*+}$ and the first measurement of
the same quantities for inclusive $B$ decay to $D^{*0}$.             
Here $B$ is the mixture of $B_d$ and $B_u$ from $\Upsilon(4S)$ decay.  
Furthermore, since more than one charm particle (or antiparticle) of the same
kind can be produced in $B$ decay, here ``inclusive $B$ branching
ratio'' is used to mean the average number of charm particles and their
antiparticles of a certain species produced in $B$ decay.  
We obtain the following results (the first error is statistical, the second
systematic of this analysis, the third is propagated from other
measurements):
${\cal B}(B\to D^0 X)    = (0.636\pm 0.014\pm 0.019\pm 0.018),
 {\cal B}(B\to D^+ X)    = (0.235\pm 0.009\pm 0.009\pm 0.024),
 {\cal B}(B\to D^{*0} X) = (0.247\pm 0.012\pm 0.018\pm 0.018),
 {\cal B}(B\to D^{*+} X) = (0.239\pm 0.011\pm 0.014\pm 0.009)$.
  The following ratio of branching ratios is not affected by most of the
systematic errors:
${\cal B}(B\to D^{*0} X)/{\cal B}(B\to D^{*+} X) =
(1.03\pm 0.07\pm 0.09\pm 0.08).$ 
We also report the first measurement of the momentum-dependent $D^{*0}$
polarization and a new measurement of the $D^{*+}$ polarization in inclusive
$B$ decay.  Using these measurements and other CLEO results and making some
additional assumptions, we calculate the average number of $c$ and $\bar c$
quarks produced in $B$ decay to be $\langle n_c\rangle\ = 1.10\pm 0.05$. 
\end{abstract}


\newpage
\renewcommand{\thefootnote}{\alph{footnote}}
\begin{center}
L.~Gibbons,$^{1}$ S.~D.~Johnson,$^{1}$ Y.~Kwon,$^{1}$
S.~Roberts,$^{1}$ E.~H.~Thorndike,$^{1}$
C.~P.~Jessop,$^{2}$ K.~Lingel,$^{2}$ H.~Marsiske,$^{2}$
M.~L.~Perl,$^{2}$ S.~F.~Schaffner,$^{2}$
% ...
\end{center}
 
\small
\begin{center}
$^{1}${University of Rochester, Rochester, New York 14627}\\
$^{2}${Stanford Linear Accelerator Center, Stanford University, Stanford,
California 94309}\\
% ...
\end{center}

\renewcommand{\thefootnote}{\arabic{footnote}} 
\setcounter{footnote}{0}

\newpage
\section{Introduction}
Measurements of weak decays of $B$ mesons are essential to testing and
understanding the standard model and determining the fundamental
flavor-mixing parameters. These measurements also provide a unique
opportunity to examine the short distance behaviour of weak decays\cite{SLS}.
Due to their large energy release, long distance corrections are expected to
be less significant in hadronic $B$ decay than in hadronic decays of charm
and strange quarks. However, the formation of hadrons in the final state is
still poorly understood and is an obstacle to predicting rates and spectra
for hadronic decays\cite{FWD,WWU}.

For several years the small observed value of the $B$ semileptonic decay
branching fraction, ${\cal B}_{SL}(B)$, seemed to disagree with theoretical
calculations \cite{SVsemi,MW,Mannel,tau,SVrare,BS}.  Recently, higher order
perturbative calculations, taking also into account the charm quark mass
\cite{BAG1,NEUB}, come close to reconciling theory
with experiment in this respect.  However, the low value of ${\cal
B}_{SL}(B)$ implies a larger than naively expected non-leptonic $B$ decay
rate \cite{HITO}.  Two mechanisms have been proposed and discussed in the
literature \cite{FWD,BAG1,NEUB,HITO,BLOK,SIMU}: an enhancement of $b\to c\bar
c s$ or of $b\to c\bar u d$.  The former would give a larger average number
of $c$ and $\bar c$ quarks, $\langle n_c\rangle$, per $B$ decay.  Since $B\to D^0$ and
$B\to D^+$ transitions (where the $D$ can also be the decay product of charm
resonance) account for a large fraction of the charm quarks produced in $B$
decay, it is important to measure accurately the branching fractions for
these transitions.  The shape of the momentum spectrum can also be compared
to models of hadronic $B$ decay and is sensitive to new production
mechanisms\cite{WWU,HITO}.  In addition, the $D$ and $D^*$ inclusive spectra
are of interest to future $B$ experiments and high energy colliders as Monte
Carlo simulations must be constrained to agree with the observed production
at the $\Upsilon(4S)$.

In this paper, we describe high statistics measurements of the differential
and total branching ratios for the inclusive $B$ decay to $D$ and $D^*$
mesons, including the first measurement of ${\cal B}(B\to D^{*0} X)$. 
For previous measurements, see references \cite{INEX,ARGINC}.

We have also measured the $D^{*+}$ and $D^{*0}$ polarizations as a function
of the $D^*$ momentum.  Since there is no complete reconstruction of the $B$
final state, we do not distinguish between $B^0$ and $B^+$.  We produce
$B\bar{B}$ states from $\Upsilon (4S)$ decays, and our generic $B$ is about
an even \cite{BPBO} admixture of $B_d$ and $B_u$ mesons and their
antiparticles . 
In the following, reference to the charge conjugate states is implicit unless
explicitly stated.

It is possible for a $B$ meson to decay to a final state containing
two $D$ mesons.  By ``inclusive $B$ decay branching ratios'' to a given $D$
species measured in this analysis we mean the``average number of $D$ and
$\bar D$'' per $B$ decay.

After a brief description of the CLEO II apparatus and data sample used, we
describe our analysis procedure in Sect.III. In Sect.IV-VI we give the
$B$ branching ratios and momentum spectra results and analysis details
specific to the four individual channels.  In Sect.VII, we show the results
on $D^*$ polarization and in Sect.VIII we discuss and summarize our results.

\section{Detector and Data Sample}
\label{sec-DET-DAT}
The data sample used in this inclusive $B$ decay analysis was taken at CESR,
the Cornell Electron Storage Ring, during 1990 -- 1994.  The sample consists
of about 2,020 $ pb^{-1}$ of integrated luminosity of $e^+ e^-$ annihilation
data taken near the peak of the $\Upsilon(4S)$ resonance and about 959
$pb^{-1}$ just below the open bottom threshold (referred to in this paper as
``continuum'' data).  The data used correspond to about 2.166 million
$B\bar{B}$ events.  We estimate the actual number of $B\bar{B}$ events by
subtracting the number of events in the continuum data after scaling by the
ratio of the luminosities and correcting for the center-of-mass energy
dependence of the continuum annihilation cross-section. We also perform minor
corrections due to minor differences in CESR running conditions at the two
energies.  This is equivalent to multiplying the integrated luminosity
collected near the peak of the $\Upsilon(4S)$ resonance by the peak cross
section $\sigma(e^+e^- \to\Upsilon(4S)) = (1.072\pm 0.019)~nb$ extracted from
our own data.  For the purpose of this analysis we assume that the
$\Upsilon(4S)$ decays exclusively to $B\bar B$.

In the CLEO~II detector \cite{DET} charged particles are tracked in a
1.5~Tesla magnetic field through three nested coaxial cylindrical drift
chambers covering $94\%$ of the solid angle.  The innermost chamber is a
six-layer straw-tube vertex detector of inner radius $4.5~cm$ with $50~\mu m$
position accuracy in the $r-\phi$ plane.  It is followed by a ten-layer
pressurized inner drift chamber with a position accuracy of $100~\mu m$ in
$r-\phi$.  The main cylindrical drift chamber \cite{DRII} contains 51 anode
layers, 11 of which are strung at angles to the z-axis progressing from about
$\pm4^\circ$ to about $7^\circ$.
It has a position accuracy of $120~\mu m$ in $r-\phi$, it gives a transverse
momentum resolution of $(\delta p_t/p_t)^2 = (0.0015 p_t)^2 + (0.005)^2$
(with $p_t$ in GeV) and a $dE/dx$ resolution (measured on Bhabha scattering
events) of $6.5\%$ for particle identification, giving good $\pi/K$
separation up to $700~MeV/c$.  The outer radius of the main drift chamber is
$1~m$.  Cathode layers are located at the inner and outer radii of the
ten-layer inner drift chamber, and at the inner and outer radii of the main
drift chamber, to improve information about the z coordinates (along the
beam) of the tracks.  Time of flight counters with $154~ps$ resolution are
located outside the drift chambers and provide additional information for
particle identification (not used in this analysis). 

Photon and $\pi^0$ detection as well as electron identification use the CsI
electromagnetic shower detector \cite{CSI}.  It consists of 7800 CsI(Tl)
crystals between the time of flight counters and the super-conducting magnet
coil in the barrel region and, in endcaps, between the drift chamber plates
and the magnet pole pieces, altogether covering 95\% of the solid angle. 

The material in the drift chamber endplates, electronics and cables
degrades the performance of the calorimeter in the endcaps, especially at the
two ends of the barrel region.  Photon candidates are restricted     
to lie in the region of the calorimeter covering the angular
region $\vert\cos\theta\vert<0.707$.  The energy calibration makes use of
Bhabha scattering and radiative Bhabhas as well as $e^+e^-\to \gamma\gamma$
reactions and $\pi^0\to\gamma\gamma$ decays \cite{MORR}.  For low
multiplicity final states, the energy resolution in the barrel portion of the
calorimeter is given by ${\sigma_E\over E}(\%) = {0.35 / E^{0.75}} + 1.9 -
0.1E$, where $E$ is the photon energy in GeV.

Muons can be identified by their penetration in the three $36~cm$ thick slabs
of iron that surround the super-conducting coil in an octagonal geometry, and
in the iron pole pieces of the magnet \cite{MUii}.

\subsection{Monte Carlo simulation}
\label{MCSIM}
To estimate detection efficiencies we generated Monte Carlo events using the
Jetset 7.3 \cite{SJO} package for continuum annihilation events and the CLEO
model for $\Upsilon(4S)\to B\bar{B}$ decays \cite{DDK}.  Separate Monte Carlo
datasets were generated for the analyses of $B\to D^0 X$ and of $B\to D^+
X$ decays.  In the $D^+$ case, the $D^+$ is allowed to decay only into the
final state we use, $K^-\pi^+\pi^+$, while the $D^-$ is allowed to decay
according to measured branching fractions of its decay modes.  About 110,000
$B\bar{B}$ events and about 35,000 continuum events that contain a $D^{+}$
were generated. An analogous procedure is followed in generating $D^0\to
K^-\pi^+$, except that the $D^0$ is also allowed to decay into $K^+K^-$ and
$\pi^+\pi^-$ final states, according to their measured decay branching
fractions relative to the $K^-\pi^+$ mode.  About 170,000 $B\bar{B}$ events
and about 150,000 continuum events that contain a $D^{0}$ were generated.
The events are then processed through a GEANT-based \cite{GEANT} simulation
of the CLEO II detector and reconstructed and analyzed as real data.  We call
these Monte Carlo datasets the ``dedicated'' Monte Carlo.

We also used a set of Monte Carlo events produced in a similiar manner,
but with all $D$ mesons decaying according to a model which incorporates the
current knowledge of their decay modes.  This statistically independent
``generic'' Monte Carlo (in which the ``true'' value of the quantities we aim
to measure is known {\it a priori}) was used to check the analysis
procedure. This type of consistency check, of course, does not exclude the
possibility of a systematic flaw that is common to both the ``dedicated'' and
``generic'' Monte Carlo data samples.

\label{MCTAG}             
The ``generic'' Monte Carlo was also used for a different purpose.  The
reconstructed tracks were associated with the simulated particles that
generated them.  We could thus generate high statistics, background free
distributions of the signals and distributions of various backgrounds.  We
called these distributions Monte Carlo Tagged (MC-tag for short)
distributions.

\section{Analysis Procedure}

We reconstruct the $D$ and $D^*$ mesons using the following exclusive decay
modes that have the largest signal to background ratios and for which the
Monte Carlo can reliably estimate the reconstruction efficiency:
\begin{eqnarray}
D^0 & \rightarrow & K^- \pi^+ \\
D^+&\to& K^- \pi^+\pi^+ \\
D^{*0}&\to&D^0\pi^0\to(K^-\pi^+)\pi^0 \\
D^{*+}&\to&D^0\pi^+\to(K^-\pi^+)\pi^+ \\
D^{*+}&\to&D^+\pi^0\to(K^-\pi^+\pi^+)\pi^0
\end{eqnarray}
In this section, we describe the selection criteria and general procedures
used in the analysis. In the sections which follow the details of the
procedures for decay modes of the $D^0$, $D^+$, $D^{*0}$ and 
$D^{*+}$ will be given.
\subsection{Selection Criteria}
\label{sec:selcrit}
The continuum background is suppressed by excluding candidates from
jet-like events. This is accomplished by using the Fox-Wolfram parameters
\cite{FOX}.  We require $R_2 \equiv H_2/H_0 < 0.5$.  This cut has an
efficiency of $99.38 \pm 0.02\%$ while rejecting $29.0 \pm 0.2\%$ of
continuum events in the $B$ decay kinematical region.  The efficiency for
this requirement is determined from our Monte Carlo simulation and agrees
with the estimate derived from the data.

Each charged track used to reconstruct a $D$ or $D^*$ is required to be
consistent with originating from the primary vertex.  If the momentum of a
track is greater than 0.250 GeV, we require that the $z$ coordinate (along
the beam) of the point of closest approach of the track to the beam-line be
within $3~cm$ of the $z$ coordinate of the event vertex and that the track's
impact parameter with respect to the beam line be less than $ 5~mm$.
For tracks with momentum less than 0.250 GeV these requirements are loosened
to $5~cm$ and $10~mm$ respectively.

Particle identification requirements were imposed based on the specific
ionization (dE/dx) measurements for the track, provided that more than 10 hits
were recorded in the main drift chamber.  The observed dE/dx had to be
within three standard deviations of that expected for the particle species
considered.

Each photon candidate shower is required to lie within the good barrel region
($\cos\theta<0.707$) of the crystal calorimeter and to have a minimum energy
of 30 MeV.  Photon candidates are also required to be well separated from
the extrapolated position of all charged tracks,
and the lateral shape of the shower should be consistent with that 
expected from an electromagnetic shower.

Candidate $\pi^0$ mesons are reconstructed from pairs of photon candidates.
If the effective mass of two photons is less than 2.58$\sigma$ away from the
expected $\pi^0$ mass, the combination is accepted as a $\pi^0$ candidate and
then is kinematically fitted to the nominal $\pi^0$ mass.
\subsection{Common procedure}
\begin{figure}[htbp]
\centerline{\psfig{file=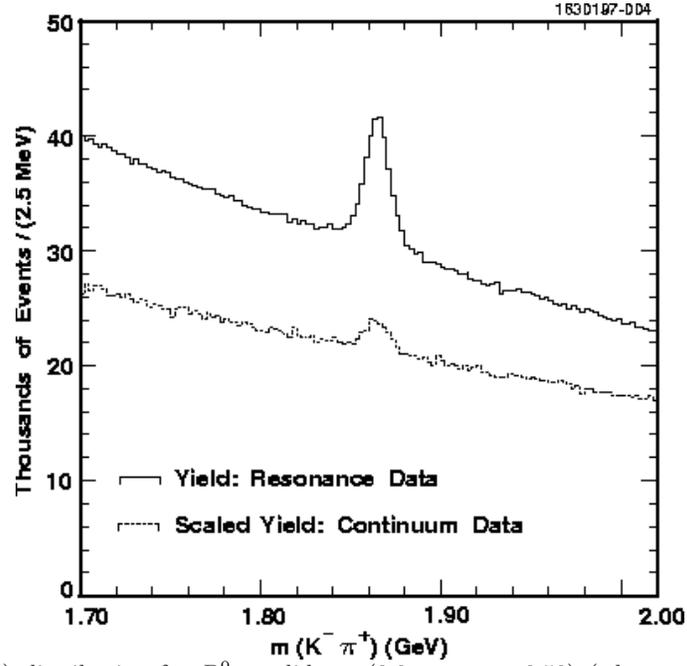,width=3.5in}}
\caption{\label{DZGLOB} The $m(K^-\pi^+)$ distribution for $D^0$ candidates
$(0.0<x_D<0.50)$ (where $x_D = p_D/p_{max}$ ) from ``on resonance'' data
($p_{max}=4.950~GeV/c$) and from ``continuum'' data ($p_{max}=4.920~GeV/c$).
The ``continuum'' distribution is scaled by the luminosity and cross section
factor 2.080.}
\end{figure}

The inclusive $B$ decay spectra and branching fractions are obtained by the
subtraction of the candidate mass distribution below $B\bar{B}$ threshold
(scaled by the ratio of luminosities and of the $e^+e^-$ annihilation cross
section) from the candidate mass distribution on the $\Upsilon(4S)$
resonance.  To illustrate the effect of continuum subtraction we show in
Fig.~\ref{DZGLOB} the $(K^-\pi^+)$ effective mass distribution of $D^0$
candidates in the sample taken at the $\Upsilon(4S)$ resonance, in the 
whole momentum interval allowed in $B$ decay, and the corresponding
distribution from the sample taken below $B\bar B$ threshold, scaled by a
factor of 2.08, the combined ratio of luminosities and of the $e^+e^-$
(non $b\bar b$) annihilation cross sections at the two energies.

In order to perform such subtraction it is convenient to use a scaling
variable.  We use the scaling variable $x$, defined as $x = p/p_{max}$ where
$p_{max} = 4.950~GeV/c$, the momentum for a $D^0$ produced in the reaction
$e^+e^-\to D^0 \overline D^0$ at a center of mass energy of $10.58~GeV$
($p_{max} = 4.920~GeV/c$ for the continuum sample, at a center of mass energy
of $10.520~GeV$).

The momenta of the charmed mesons are measured in the $\Upsilon(4S)$ rest
frame rather than in the $B$ rest frame.  Since the $\Upsilon(4S)$ mass
(10.58 GeV) is slightly above the threshold for B meson pair production
(10.56 GeV), the $B$ mesons are not at rest. The $B$ momentum ranges from
about 265 MeV to 355 MeV ($\pm$ one standard deviation).  This motion smears
the value of $x$ relative to what it would be if the $B$ were at rest.  A
Monte Carlo study shows that the smearing in the variable $x$ varies from
$\pm 0.013$ to a maximum of $\pm 0.020$.
Taking this effect into account, the maximum value of $x$ for $B$ decay to
$D$ is 0.506, and 0.496 for decay to $D^*$.
\subsubsection{Spectra and the $B$ branching fractions}
In order to measure the spectra, we divided our sample of charmed particle
candidates into 10 $x$ bins (20 for the $D^0$ where the statistics
are high).  For each $x$ bin we
generated the effective mass distributions of the $D$ candidates (already
selected as candidate $D^*$ decay products in the case of $D^*$, see below)
from ``on resonance'' and ``continuum'' data and performed bin-by-bin the
scaled continuum subtraction, obtaining the mass distributions of the
candidates from $B$ decay.  We then fitted this mass distribution, in each
$x$ bin, to the sum of the $D$ signal and the various backgrounds.
The shape of the signal, its parametrization and the different backgrounds
will be described later.

We performed an identical analysis (except for the non-existent continuum
subtraction) on the Monte Carlo simulated events, to find the detection
efficiencies as a function of $x$.
The inclusive $B$ decay branching ratios are calculated from the integral of
the appropriate $D$ (or $D^*$) efficiency-corrected spectrum.

\subsubsection{$D^*$-tagging and ``background-free'' $D$ samples}
\label{DSTAG}
\begin{figure}[htbp]
\centerline{\psfig{file=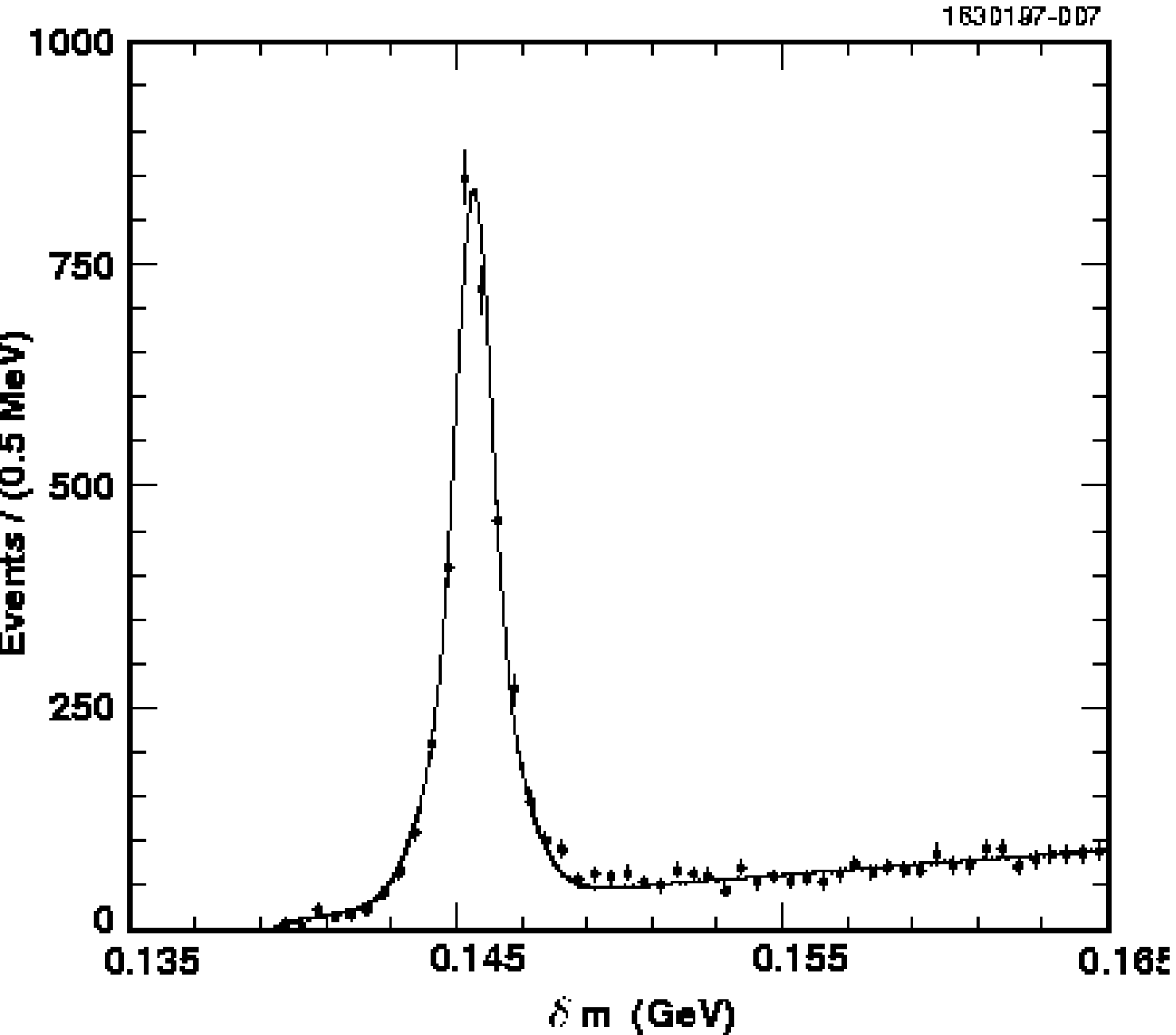,width=3.5in}}
\caption[]{The $\delta m$ distribution for $D^{*+}\rightarrow D^0\pi^+$
candidates (data).}
\label{DELMFPA} 
\end{figure}

\begin{figure}[htbp]
\centerline{\psfig{file=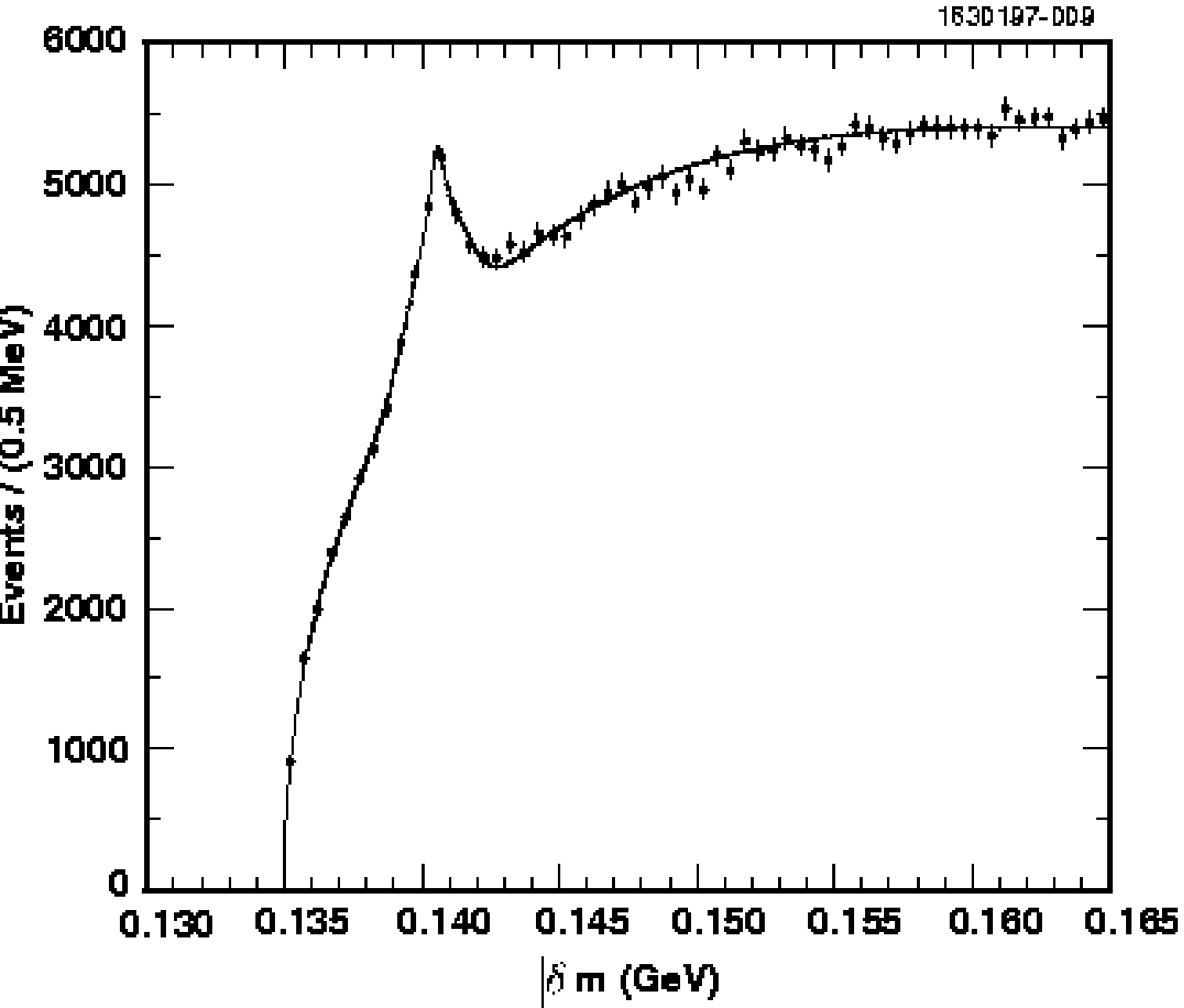,width=3.5in}}
\caption[]{The $\delta m$ distribution for $D^{*+} \rightarrow D^{+} \pi^{0}$
candidates (data).}
\label{DELMFPC} 
\end{figure}

\begin{figure}[htbp]
\centerline{\psfig{file=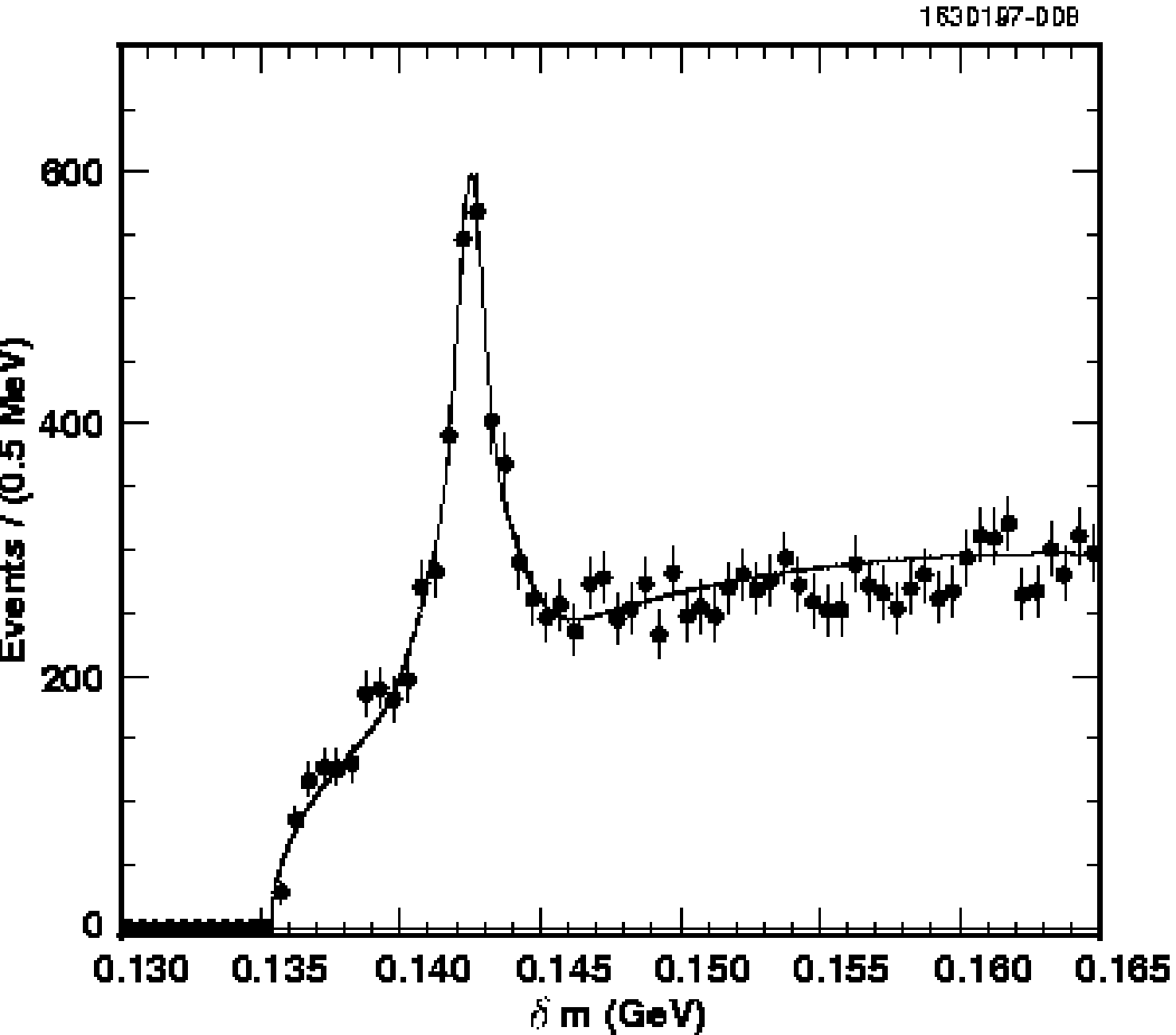,width=3.5in}}
\caption{\label{DELMFPB} The $\delta m$ distribution for $D^{*0}\rightarrow
D^0\pi^0$ candidates (data).}
\end{figure}

We have studied the shape of the $D$ signal and its momentum dependence, in
data and in Monte Carlo simulation, using $D^0$s from $D^{*+}\to D^0\pi^+$
decays and $D^+$s from $D^{*+}\to D^+\pi^0$ decays.  The $D^0$
sample selected this way is of known flavor because of the charge of the
$D^{*+}$. It was also used to obtain the shape and momentum dependence of the
``switched-mass'' $D^0$ background discussed later in Sect.~\ref{SWMAS}.  We
obtain a very low background sample of $D^0$s from $D^{*+}$ by selecting
events in a $\pm 2\sigma$ region around the peak of the mass difference
$\delta m\equiv m(K^-\pi_1^+\pi_2^+) - m(K^-\pi_1^+)$ distribution ($\sigma
\sim 0.8$ MeV).  We obtain a low background sample of $D^+$s
using a similar selection on the $\delta m\equiv m(K^-\pi^+\pi^+\pi^0)
-m(K^-\pi^+\pi^+)$ distribution ($\sigma \sim 1.0$ MeV).  The $\delta m$
distributions for $D$ candidates in the $D$ mass signal region in the three
channels considered in this analysis are shown in
Figs.~\ref{DELMFPA},\ref{DELMFPC},\ref{DELMFPB}(data).  Most of the residual
background in the $D^*$-tagged $D$ mass distribution is eliminated by
subtraction of the $\delta m$ sidebands.

Due to the fact that the signal in $\delta m$ is very close to threshold, it
is difficult to choose sidebands that are wide enough to balance the number
of background events under the $\delta m$ peak and far enough from the peak.
Our choices for signal and sideband regions are shown in
Table~\ref{tab:dm}.  The signal region has a half width, and the sidebands
a total width, of approximately twice the Gaussian $\sigma$ of the $\delta
m$\ signal.  The position of the signal peak is obtained from the data.

\begin{figure}[htbp]
\centerline{\psfig{file=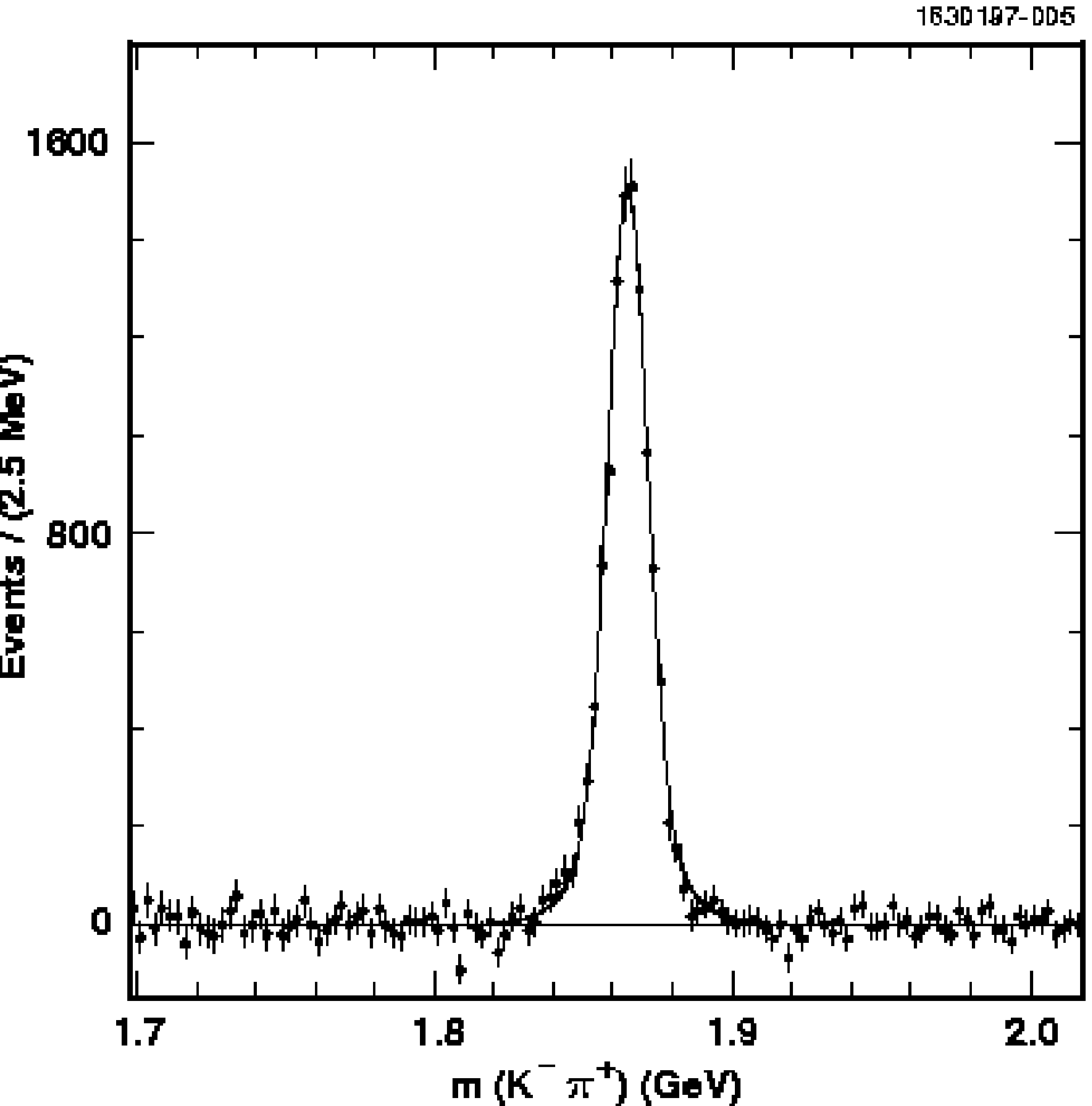,width=3.5in}}
\caption[]{The ``background-free''  $m(K^-\pi^+)$ distribution for
$D^0$ mesons that are decay products of $D^{*+}\to D^0\pi^+$.
$0.0<x_D<0.50$. The two-Gaussian fit of the distribution is also shown.} 
\label{DZSH} 
\end{figure}

\begin{figure}[htbp]
\centerline{\psfig{file=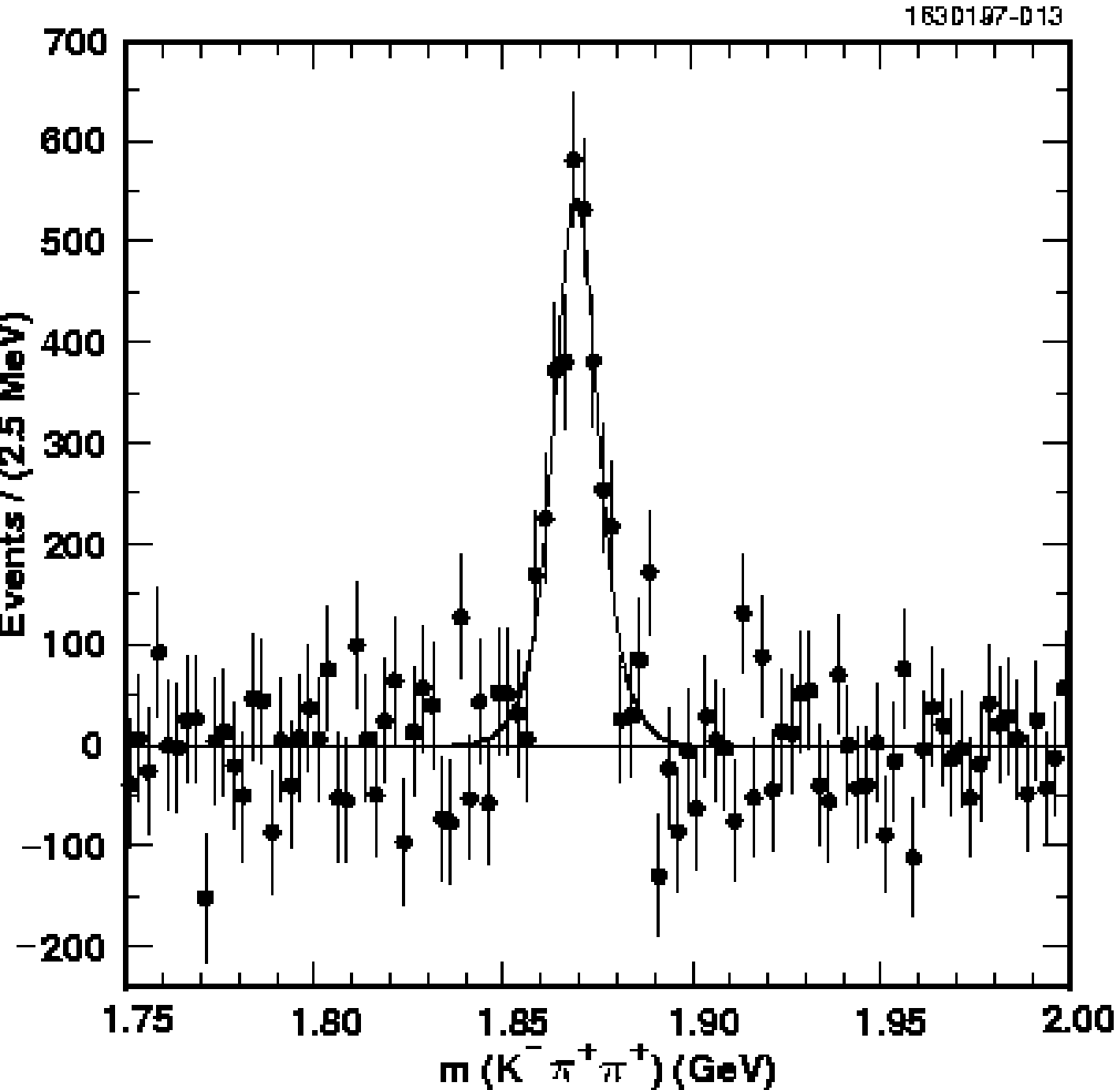,width=3.5in}}
\caption[]{
The ``background-free'' $m(K^-\pi^+\pi^+)$ distribution for $D^+$ mesons that
are decay products of $D^{*+}\to  D^+\pi^0$. $ 0.0<x_D<0.50$.  The two-Gaussian
fit of the distribution is also shown.} 
\label{DPSH} 
\end{figure}

Before subtraction, the number of events in the $\delta m$ sidebands must be
scaled to the estimated number of background events in the peak region.  Two
methods were used to estimate the scaling factor.  The first method is to fit
the $\delta m$ distribution with a smooth threshold function background and a
double-Gaussian signal and calculate the scale factor by integration of the
background function.  The second method is to obtain the scale factor by
fitting the $D$ invariant mass distribution in the $\delta m$\ peak and side
band regions with a two-Gaussian $D$ signal function and a polynomial
background.  The scale factor is then given by the ratio of the background
levels under the $D$ peaks.  Fig.~\ref{DZSH} was obtained by the second
method.  The results of the two methods agree and the scale factors are close
to 1.1.  The same procedure was also used for the analysis of the Monte Carlo
sample.  The candidate $D$ mass distributions obtained from data by this
$D^*$-tagging procedure are shown in Figs.~\ref{DZSH} and \ref{DPSH}.

\subsection{Estimate of the detection efficiency}
The detection efficiencies for the charm mesons studied were obtained analyzing
the dedicated Monte Carlo simulation samples described in Sect.~\ref{MCSIM}
with the same procedures and selections used to analyze data.  However,
measurement errors are slightly underestimated in the simulation.  In practice
the only relevant parameter is the width of the reconstructed mass.
Consequently, we analyzed the data using the parameters extracted from the data
themselves, and analyzed the Monte Carlo samples using the parameters
extracted from the Monte Carlo samples.

\subsection{Calculation of the $B$ decay branching fractions}
The differential $B$ decay branching fractions are 
calculated bin-by-bin from the
following equations:  
\begin{equation}
   {d\cal B}( B \rightarrow D X )\times{\cal B}(D) 
   = \frac{dN_{D}}{ 2 N_{B\bar{B}}\times\epsilon} \label{BRA}
\end{equation}
\begin{equation}
   {d\cal B}( B \rightarrow D^{*} X )\times{\cal B}(D^{*})\times{\cal B}(D) 
   = \frac{dN_{D^{*}}}{ 2 N_{B\bar{B}} \times \epsilon } \label{BRB}
\end{equation}
where $dN_{D}$ ($dN_{D^{*}}$) is the yield of $D$ ($D^{*}$) in that bin,
$ N_{B\bar{B}} $ is the number of of $e^+e^- \to B\bar{B}$ events 
produced and $\epsilon$ is the $x$-dependent detection  efficiency. 
 We use CLEO results \cite{BUT,AKER,BALE}  for the $D^*$ and $D$
absolute branching fractions
\begin{eqnarray}
{\cal B}(D^{*+}\rightarrow D^0\pi^+)& = &(68.1\pm 1.0\pm 1.3)\%\label{DSPBR1} \\
{\cal B}(D^{*+}\rightarrow D^+\pi^0)& = &(30.8\pm 0.4\pm 0.8)\%\label{DSPBR2} \\
{\cal B}(D^{*0}\rightarrow D^0\pi^0)& = &(63.6\pm 2.3\pm 3.3)\%\label{DSZBR}  \\
{\cal B}(D^0\rightarrow K^-\pi^+)& = & (3.91\pm 0.08\pm 0.17)\%\label{DZBR}  \\
{\cal B}(D^+\rightarrow K^-\pi^+\pi^+)& = &(9.19\pm 0.6\pm 0.8)\%\label{DPBR}
\end{eqnarray}
The errors shown in our results for the differential spectra are relative,
{\em i.e.} bin-to-bin errors.  They do not include the error from 
smoothing the fitting function parameters and the errors of the detection
efficiency. The overall statistical error on the $B$ decay branching fraction
is, however, derived from the integral of the spectrum when no smoothing was
performed.
\subsection{Systematic Errors}
\label{ERRORS}
In this section we discuss the sytematic uncertainties of the $B\to D\ X$ and
$B\to D^*\ X$ analyses.  Uncertainties related to specific decay channels are
deferred to the relevant sections.  
We estimate a 1\% uncertainty in charged track detection efficiency (2\% for
$p(\pi)$ below about 200~MeV/c) and 5\% in $\pi^0$ detection \cite{BARISH}.

These errors are coherent for each kind of track and hence should be
multiplied by the respective number of tracks in the decay under study.
However, we obtain the $B$ decay branching fractions by dividing the product
of branching fractions (Eqs.~(\ref{BRA}) and (\ref{BRB})) by the $D$ (and $D^*$)
decay branching fractions measured in our own experiment.  Hence, to the
extent that the kinematical configurations and the data set used overlap,
most tracking errors are eliminated.  We estimate a residual systematic error
of 0.5\% per charged track (1.0\% if below 200~MeV/c) and 2\% for $\pi^0$
whenever there is compensation.

We have studied the effect on the branching fraction of the track quality,
geometry cuts and event shape cuts by succesively removing them and then
measuring the effect on the $B$ branching fractions.  The errors specific to
each final state will be quoted in the appropriate sections. 

In measuring ${\cal B}(B\to D^{*} X)$ the subtraction of the background due
to association of a true $D$ and a random $\pi$ is performed by subtracting
the $D$ candidate mass distribution of the $\delta m$\ sidebands from that
in the $\delta m$\ peak region.  The result is slightly dependent on the
choice of the width of the $\delta m$\ intervals (Table~\ref{tab:dm}).  We
varied the widths of these intervals by $\pm 0.5$~MeV in each case and
derived a relative uncertainty in the $B$ branching fraction of 0.9\% for
$B\to D^{*+}$ and of 1.2\% for $B\to D^{*0}$.

In the case of $B\to D^*$ decays, the branching fraction is also sensitive to
the value of the scale factor used in the $\delta m$\ sideband subtraction
This effect contributes a systematic error which is estimated to be the
change in the branching fraction corresponding to a $1 \sigma$ change in the
scale factor.

The value of $N_{B\bar{B}}$ in Eqs.~(\ref{BRA}) and (\ref{BRB}) is affected
by a relative error of 1.8\% (Sec.~\ref{sec-DET-DAT}).  Finally, we take into
account the statistical error on the efficiency, calculated from the Monte
Carlo simulation.  All the above errors are combined in quadrature to give
the relative systematic error on our measurements.

The third error in the $B$ branching fractions arises from the propagation of
errors in the $D$ and $D^*$ decay branching fractions, in so far as they are
not coherent with the errors in this analysis.

\section{Inclusive $B\to D^0$ Decay}

\label{DZSEC} 
\begin{figure}[htbp]
\centerline{\psfig{file=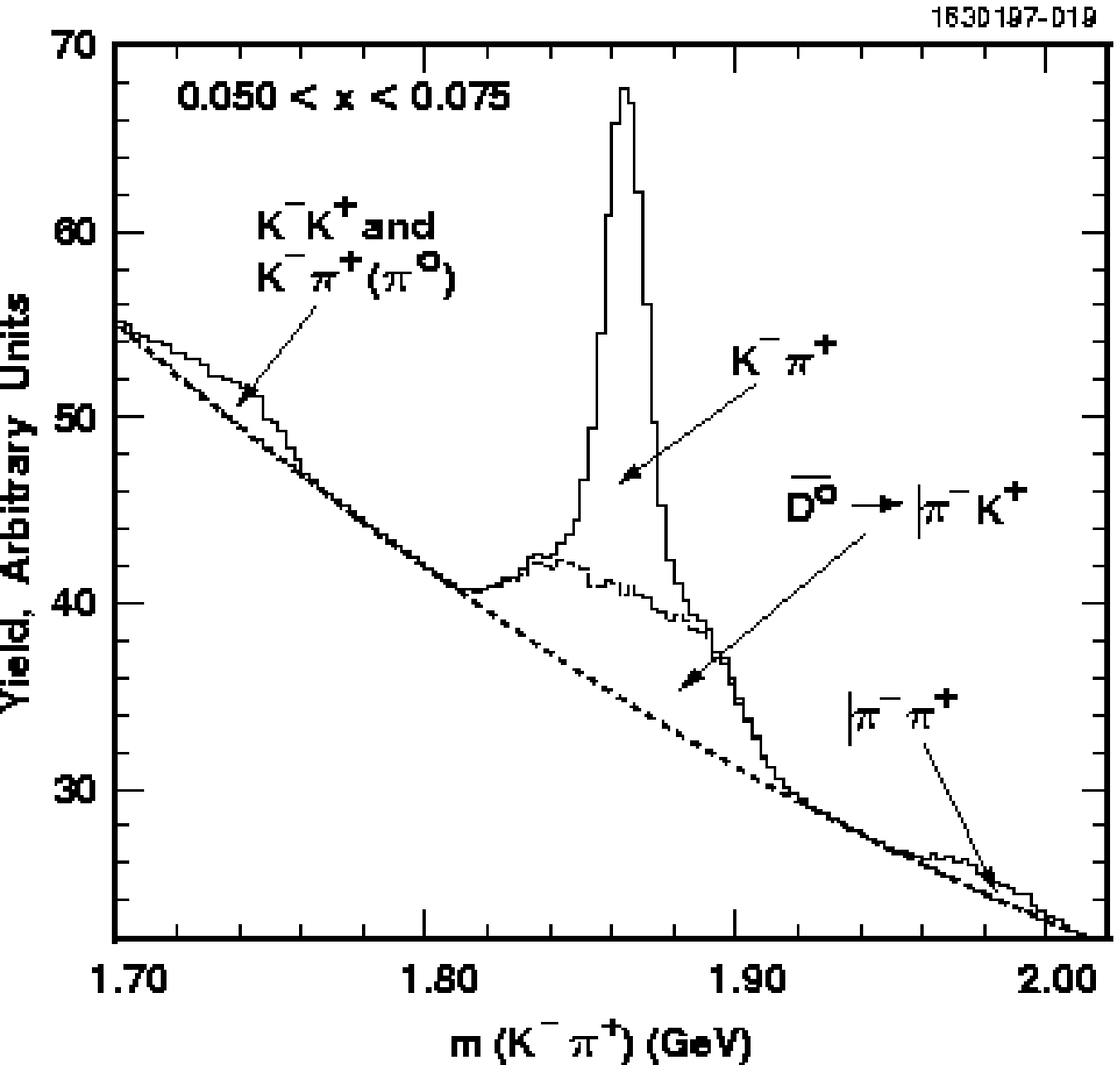,width=3.5in}}
\caption{Representation of the functions used to fit the $D^0$
signal and the various backgrounds for the momentum bin $0.050<x<0.075$.
Note the offset of the vertical scale.}
\label{fun1}
\end{figure}

\begin{figure}[htbp]
\centerline{\psfig{file=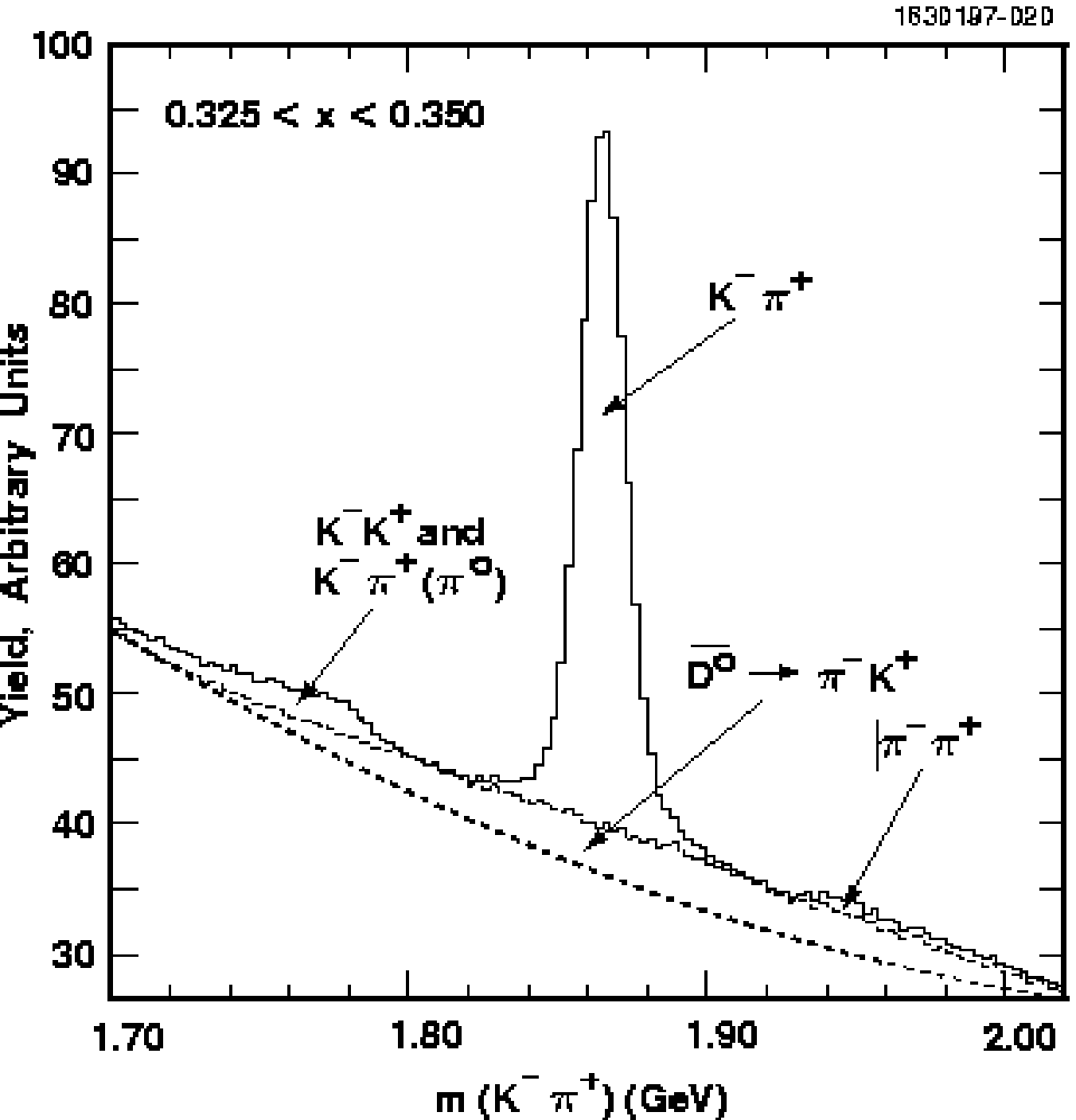,width=3.5in}}
\caption{Representation of the functions used to fit the $D^0$
signal and the various backgrounds for the momentum bin $0.325<x<0.350$. 
Note the offset of the vertical scale.}
\label{fun2}
\end{figure}

The continuum-subtracted $m(K^-\pi^+)$ distribution in any
given $D^0$ momentum interval is the sum of the $D^0$ signal plus various
backgrounds.  This is illustrated in Figs.~\ref{fun1} and \ref{fun2} which
show the $m(K^-\pi^+)$ distributions for one of the lowest momentum intervals
($0.050<x<0.075$), and for a higher momentum interval ($0.325<x<0.350$),
respectively. We see a prominent signal and three backgrounds: (i) a
combinatorial background; (ii) the ``switched mass'' background that
contributes to the signal region and (iii) two ``satellite bumps" on either
side of the signal. Backgrounds (ii) and (iii) complicate the fit to the
signal region and its immediate vicinity.  It is not possible to establish
the shape of the combinatorial background without a reliable knowledge of the
shapes and amounts of backgrounds (ii) and (iii).  A detailed discussion of
the signal shape and of these backgrounds is presented below.

\subsection{Switched mass background}
\label{SWMAS}
$\overline{D^0} \rightarrow \pi^-K^+$ 
events may be misidentified as $D^0\to K^-\pi^+$
when there is insufficient $K/ \pi$ discrimination. This background
complicates the extraction of the $D^0$ yield, particularly at low $D^0$
momenta ($p<700~MeV$) where this ``switched mass'' background distribution is
similiar in shape to the signal and peaks under it (see Fig.~\ref{fun1} and
\ref{fun2}).   For a $D^0$ at rest and without particle identification the area
of this background within 2$\sigma$\ of the signal peak would 
equal the signal area.  However, this background, within such limits,
quickly decreases to about 10\% of the area of the signal for $D^0$ momentum
above $1$ GeV and decreases further at higher momenta.  For $x_D > 0.50$, the
invariant mass distribution of these doubly misidentified events becomes so
broad that it can be absorbed in the polynomial parameterization of the
combinatorial background.  The amount of this type of background is strongly
affected by the use of particle identification.             

\begin{figure}[htbp]
\centerline{\psfig{file=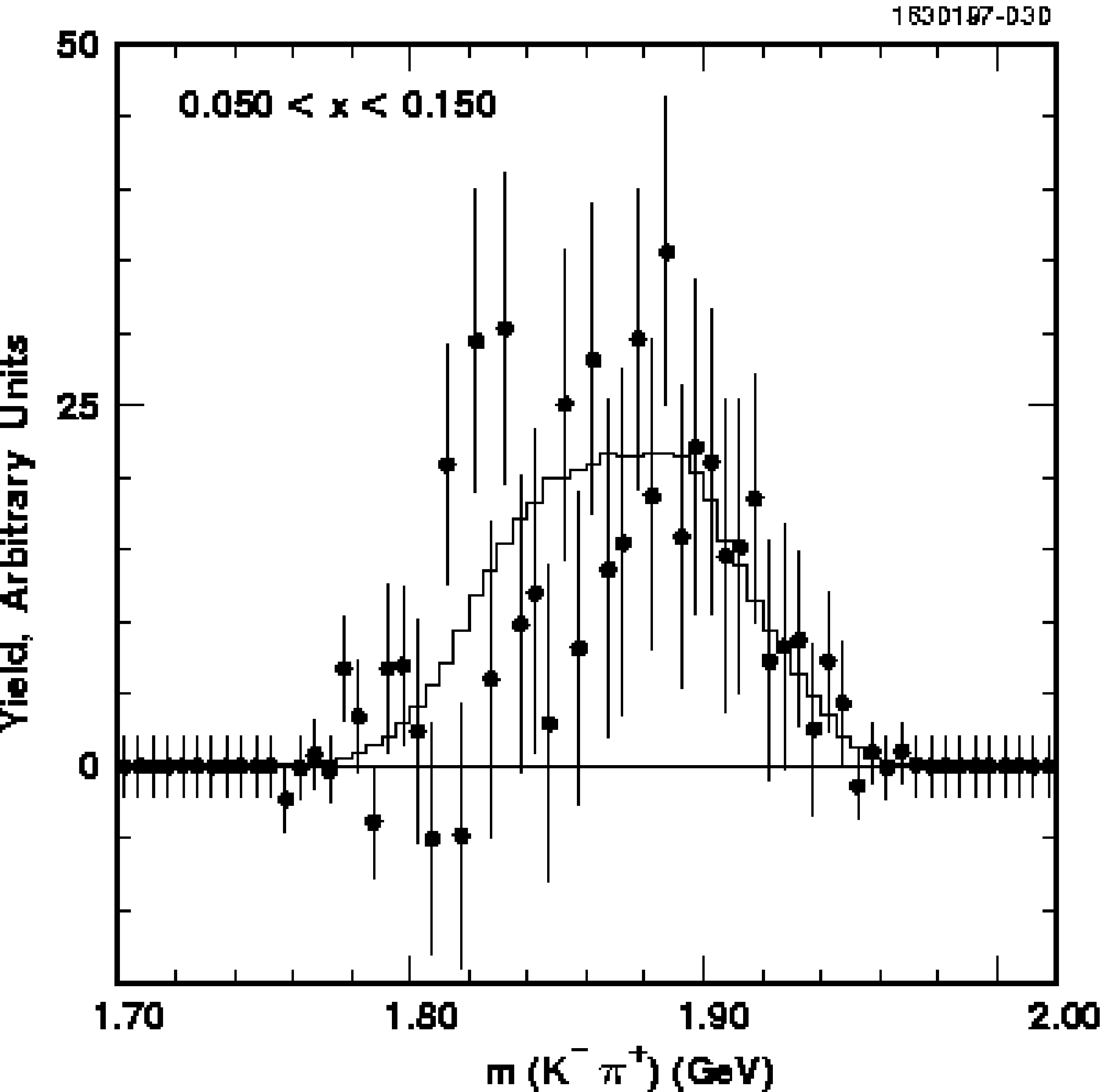,width=3.5in}}
\caption{The $m(K^-\pi^+)$ distribution for $\overline{D}^0\rightarrow K^+\pi^-$
after switching particle identity.  Points with error bars are from data, the
histogram is obtained from Monte Carlo using track tagging (MC-tag). Momentum
bin $0.05<x<0.15$.} 
\label{SW1}
\end{figure}

\begin{figure}[htbp]
\centerline{\psfig{file=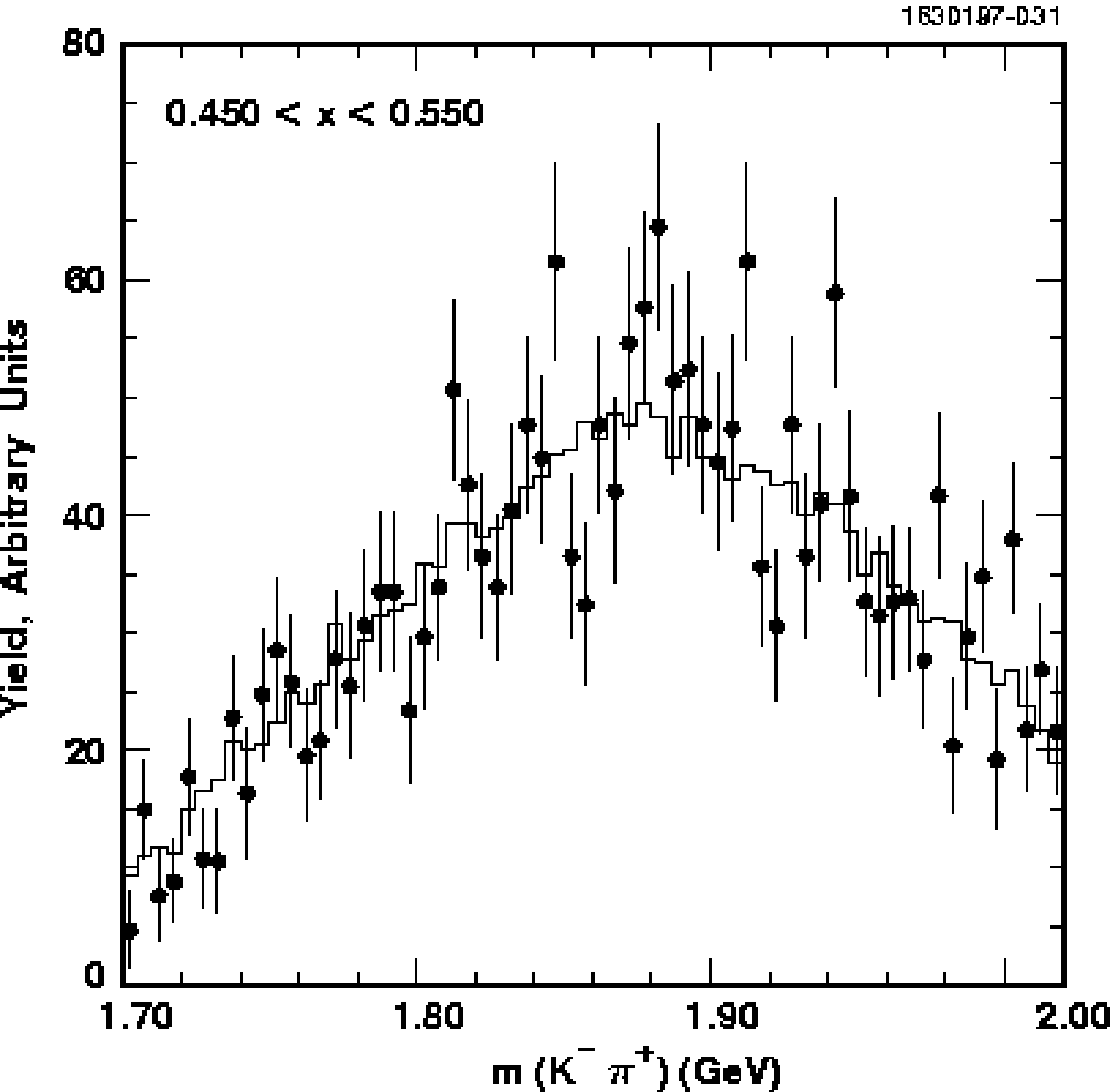,width=3.5in}}
\caption{The $m(K^-\pi^+)$ distribution for $\overline{D}^0\rightarrow K^+\pi^-$
after switching particle identity.  Points with error bars are from data, the
histogram is obtained from Monte Carlo using track tagging (MC-tag). Momentum
bin $0.45<x<0.55$.} 
\label{SW2}
\end{figure}

We studied the shape and relative size of this background and its   
dependence on $D^0$ momentum in the data and in
the Monte Carlo simulation using $D^{*+}$-tagged $D^0$s (Sec.~\ref{DSTAG}),
whose flavor is tagged by the charge of the $\pi^+$.  We produced ``switched
mass'' distributions, by interchanging the mass assignments of the $K$ and
$\pi$ tracks for each event, provided that the inverted assignements were
consistent with our particle identification criteria.  We then subtracted the
``switched mass'' distribution in the $\delta m$ sidebands from that in the
$\delta m$ peak region.  Figs.~\ref{SW1} and \ref{SW2} show the resulting
$m(\pi^-K^+)$ distributions for a low $D^0$ momentum interval, and for a
higher $D^0$ momentum interval.  We also calculated, as a function of
momentum, the area of this background relative to the area of the    
correctly reconstructed $D^0$ signal also obtained as described in
Sec.~\ref{DSTAG}. 

With our statistics it is difficult to smoothly parameterize the
shape of this background as a function of momentum.  Figs.~\ref{SW1} and
\ref{SW2} show, however, that the switched mass distributions obtained with
the MC-tag analysis of the generic Monte Carlo simulation provide an
excellent, parameter-free description of the $D^{*+}$-tagged data
distributions (Sec.~\ref{DSTAG}).

\begin{figure}[htbp]
\centerline{\psfig{file=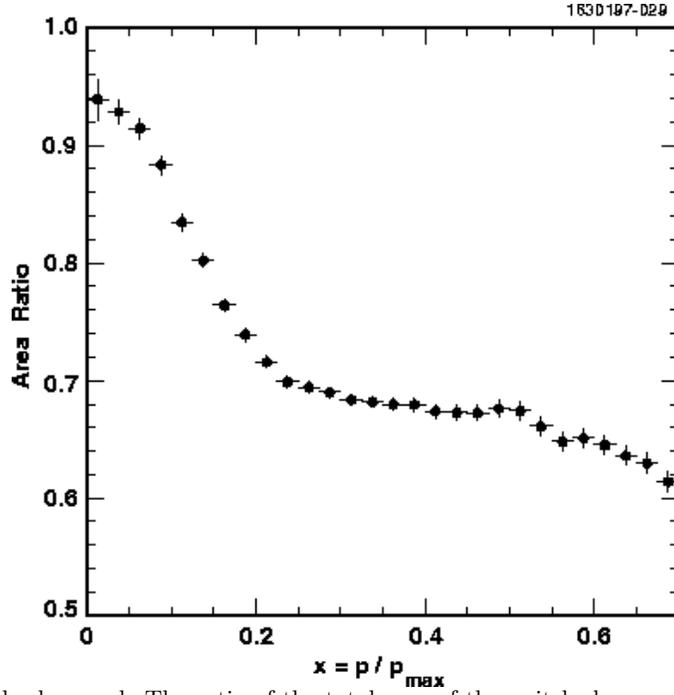,width=3.5in}}
\caption{Switched $K\pi$ background. The ratio of the total area of the
switched mass distributions (satisfying our particle identification
selections) to the signal area, vs scaled momentum $x=p/p_{max}$.}
\label{SW0}
\end{figure}

The preceding analysis is performed both on the data and on the dedicated
Monte Carlo samples.  The momentum dependence of the ratio of the switched
mass distribution area relative to the area of the signal is shown in
Fig.~\ref{SW0} as obtained from the MC-tag sample.  Similar dependences are
obtained using the $D^{*+}$-tagged samples of data and dedicated Monte Carlo.
We have checked that changing from one dependence to another has minimal
impact on the final result.

\subsection{Background from misidentified $D^0\to K^-K^+$, 
$D^0\to\pi^-\pi^+$ and $D^0\to K^-\pi^+\pi^0$ }
The Cabibbo-suppressed $D^0$ decay modes, $D^0 \to K^-K^+$ and $D^0
\to \pi^-\pi^+$, produce asymmetric enhancements on opposite sides of the
signal region when one of the tracks is misidentified as a $\pi$ or $K$,
respectively (see Fig.~\ref{fun1} and \ref{fun2}).  The $K^-\pi^+$ effective
mass distribution from the decay $D^0 \to K^-\pi^+\pi^0$ introduces an
additional small background at the lower edge of our mass spectrum.
Switching the kaon and pion masses in these decays produces an additional
background at masses well below the $D^0$ mass but with tail up to the $D^0$
mass.  If these contributions are not 
included in the shape of the background, it is possible to overestimate the
level of combinatorial background and thereby underestimate the signal.  Not
taking into account the presence of these enhancements both in fitting the
data and in fitting the dedicated Monte Carlo for obtaining the efficiency,
results in a change of the order of 1\% in the branching fraction.  
Also in this case we have found that the momentum dependence of the shape and
of the area ratio obtained by the MC-tag procedure from the generic Monte
Carlo is 
quite good and superior to that obtained using complicated analytical
parametrizations.  The size of these backgrounds relative to the signal
was taken from the Monte Carlo simulation, but we have checked that letting
their normalization float in the fit did not appreciably alter the result. 
\subsection{ Combinatorial background}
The combinatorial background, which is nearly flat in the $m(K^-\pi^+)$
region of interest, is parameterized by a second order polynomial.
\subsection{Raw $D^0$ spectrum and the shape of the signal} 
It is very important to obtain an accurate representation of the signal shape
and specifically of its tails, because of the presence of the mass-switched
background and its correlation with the signal.  We used two diferent signal
parameterizations. 
\subsubsection{Analytical parameterization}
\label{DZSHAPE} A Gaussian function does not give a sufficiently accurate
parameterization of the $D^0$ signal.  Track measuring errors are different
in different events because of the geometrical orientation of the $D^0$ decay
products           
in the detector and their overlap with other tracks.  In order to take into
account this variation of the errors, one possibility is to use a double
Gaussian with the two Gaussians constrained to have the same mean.  The
parameters of the signal shape are the mean, the width of the
narrower Gaussian, $\sigma_1$, the ratio of the widths of the wider to the
narrower Gaussian, $\sigma_2/\sigma_1$, and the ratio of the area of the
wider Gaussian to the total area, $A_2/A_{tot}$.
There are strong correlations among these three shape parameters.
It is then difficult to obtain a smooth $D^0$ momentum dependence for them if
they are allowed to float when fitting the signal in each momentum bin.

We obtained satisfactory results for these signal shape parameters by (i)
combining ``on resonance'' and ``continuum'' $D^0$ candidates together to
obtain higher statistics, especially close to the kinematic limit, or (ii) by
using the $D^*$-tagged, background-free $D^0$ signal (Sec.~\ref{DSTAG}).  We
chose the set that gave the best fit to the mass distributions and smaller
errors for the parameters.  In order to minimize the bin-to-bin statistical
fluctuations, we re-binned the $D^0$ sample in only 9 momentum
intervals $(0.025 < x < 0.475)$  and studied the position and shape of the
$D^0$ signal peak as a function of momentum.  
We smoothed the momentum dependence of the signal
shape parameters so obtained with polynomial functions of the momentum.  We
then obtained the raw $D^0$ spectrum by fitting the $m(K^-\pi^+)$
distribution in 20 $x$ bins between 0.0 and 0.5 keeping $\sigma_2/\sigma_1$
and $A_2/A_{tot}$ fixed at the smoothed values.  $\sigma_1$ is allowed to float
and varies from about 6 MeV at low momentum to about 8 MeV at the maximum
momentum; $\sigma_2/\sigma_1$ and $A_2/A_{tot}$ are approximately constant,
the first at about 2.7 and the second at about 0.15.

The same procedure was followed to analyze the events from the Monte Carlo.
The comparison of the data and Monte Carlo shape parameters shows that the
Monte Carlo simulation underestimates the track measuring errors (reflected
essentially in $\sigma_1$) by about 10\%.

\subsubsection{Use of Monte Carlo simulation with track tagging}
An alternative way of parameterizing the $D^0$ signal shape and a check of
the double-Gaussian parameterization, is to use the histograms provided by
our generic Monte Carlo simulation and tagging the reconstructed tracks
(MC-tag, Sect.~\ref{MCSIM}).  In constructing these histograms we have used
the same selection criteria used in the analysis of data and of dedicated
Monte Carlo already described.

We have just seen that our Monte Carlo simulation underestimates by about
10\% the overall width $\sigma$ of the $D^0$ signal over the whole $D^0$
spectrum. It is however likely that the Monte Carlo simulation reproduces
more accurately the dependence of $\sigma$ on track position, orientation and
overlap with other tracks, all factors that cause the non-Gaussian spread in
measuring errors seen in the data.  We then expect that the {\it shape} of
the $D^0$ signal obtained by tagging the Monte Carlo tracks will be a good
representation of the data if corrected for the overall width.  We have
expressed this correction by one parameter (then fitted on the data) that
changes the width of the signal mass distributions without otherwise altering
its shape.

We used the histograms obtained by this procedure as fitting
functions of the ``background-free'' signal distributions obtained by
$D^{\star +}$-tagging (Sect.~\ref{DSTAG}).  We obtained excellent fits,
altough not superior to the double-Gaussian fits.  We obtained $D^0$ spectra
using these distributions as fitting functions.  The $D^0$ yields differed
from those obtained with the double-Gaussian by less than a tenth of the
statistical error.
\subsection{Results}
\begin{figure}[htbp]
\centerline{\psfig{file=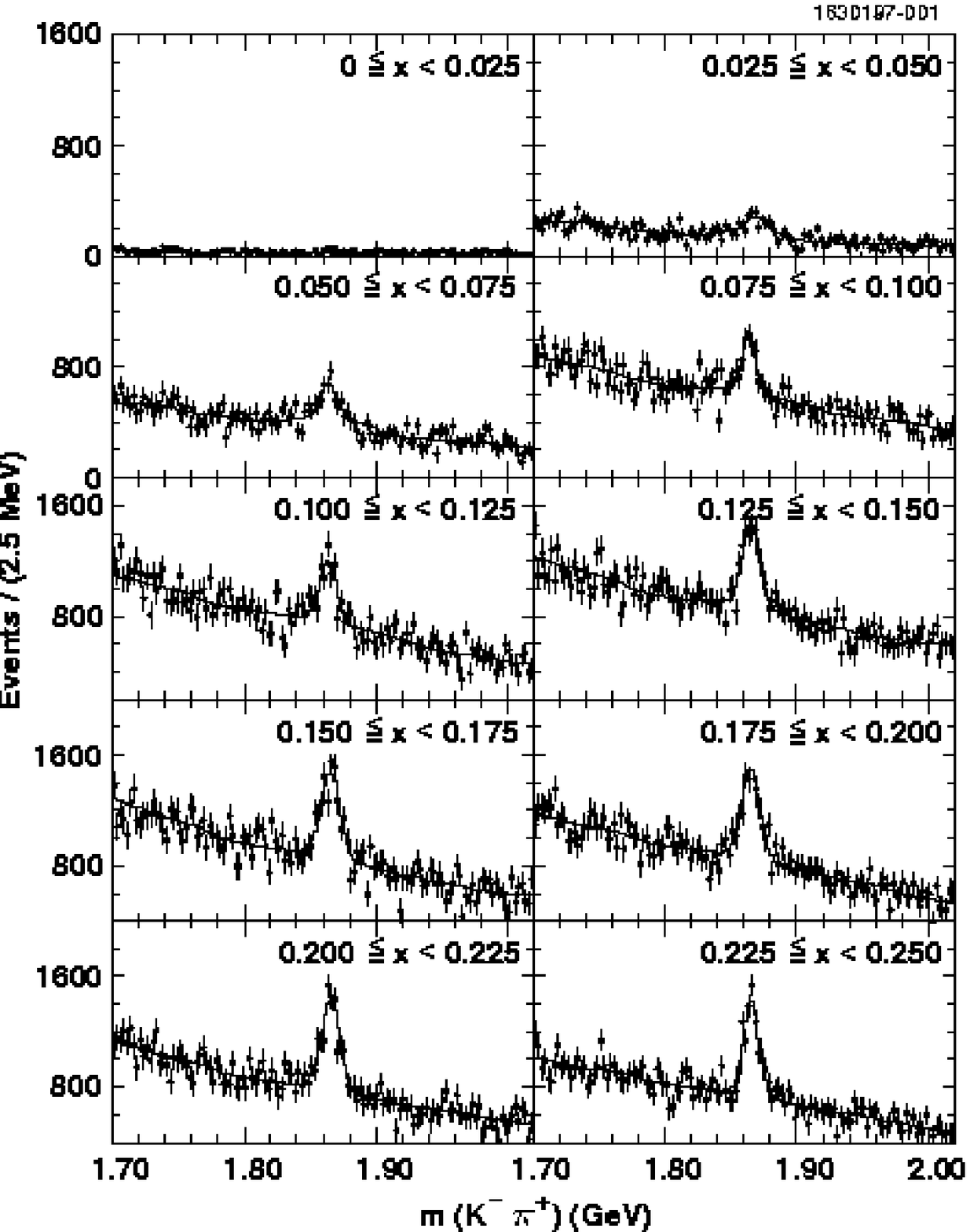,width=6.0in}}
\caption{The $m(K^-\pi^+)$ distribution for $B\to D^0\ X$ candidates in ten
$x$ bins from 0 to 0.25.  The line is the result of the fit.}
\label{figD01} 
\end{figure}

\begin{figure}[htbp]
\centerline{\psfig{file=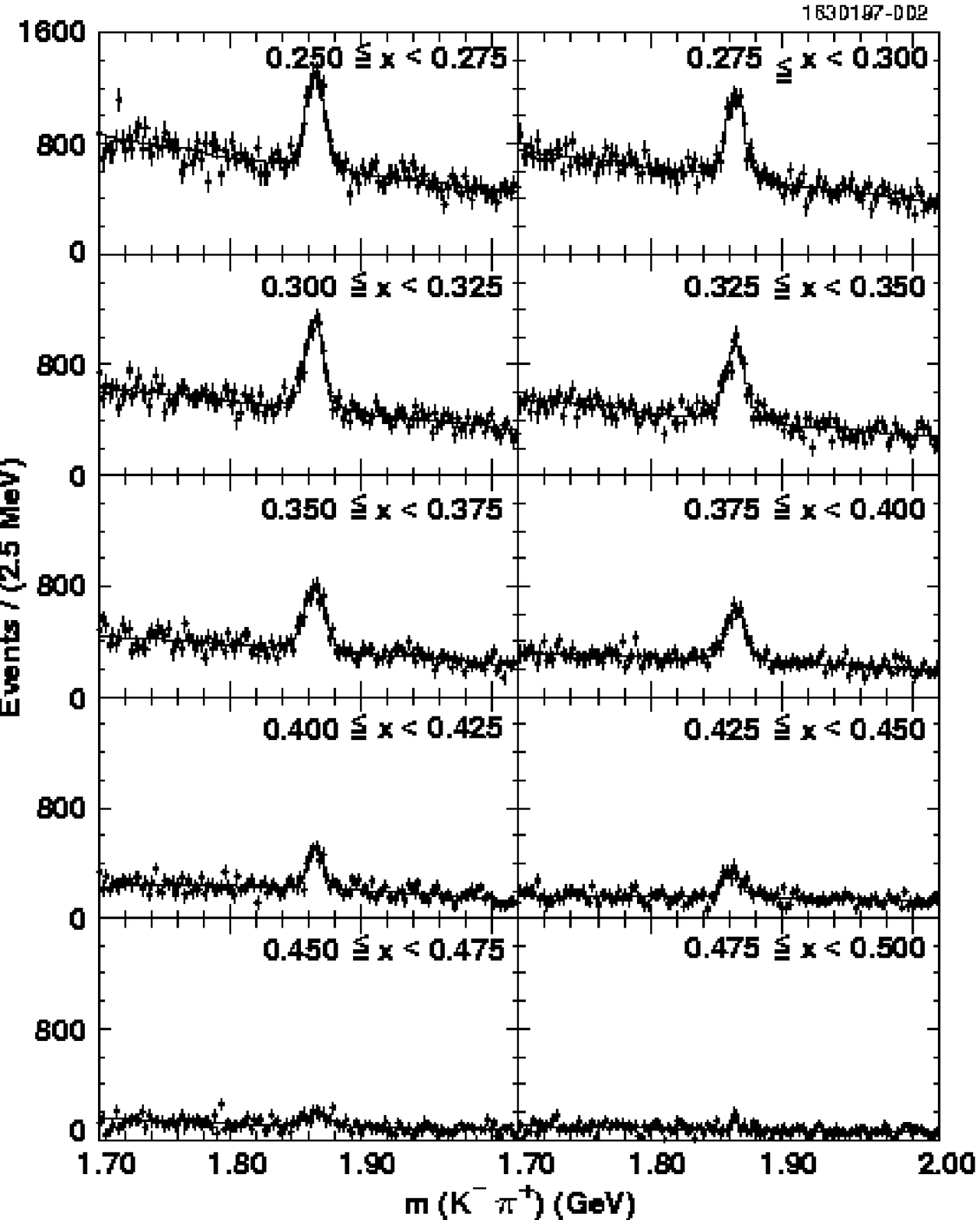,width=6.0in} }
\caption{The $m(K^-\pi^+)$ distribution for $B\to D^0\ X$ candidates in ten $x$
bins from 0.25 to 0.50.  The line is the result of the fit.} 
\label{figD02} 
\end{figure}

Particle identification (Sect.~\ref{sec:selcrit}) reduces the combinatorial
background by about a factor of 2 and improves the statistical error while
slightly increasing the systematic error due to the small uncertainty on the
accuracy of the simulation of the ionization measurements.  We performed our
analysis both with and without particle identification requirements and
obtained results that were in good agreement.  The results presented here
were obtained using particle identification and the double-Gaussian
parameterization of the signal.  In Figs.~\ref{figD01} and \ref{figD02} we
show the fitted mass distributions in the 20 $x$ bins.  Summing the spectrum
over the 20 bins for $0.0<x_D<0.5$ gave a raw yield of $62,648\pm 1,394$
$D^0$ from $B$ decay.

\begin{figure}[htbp]
\centerline{\psfig{file=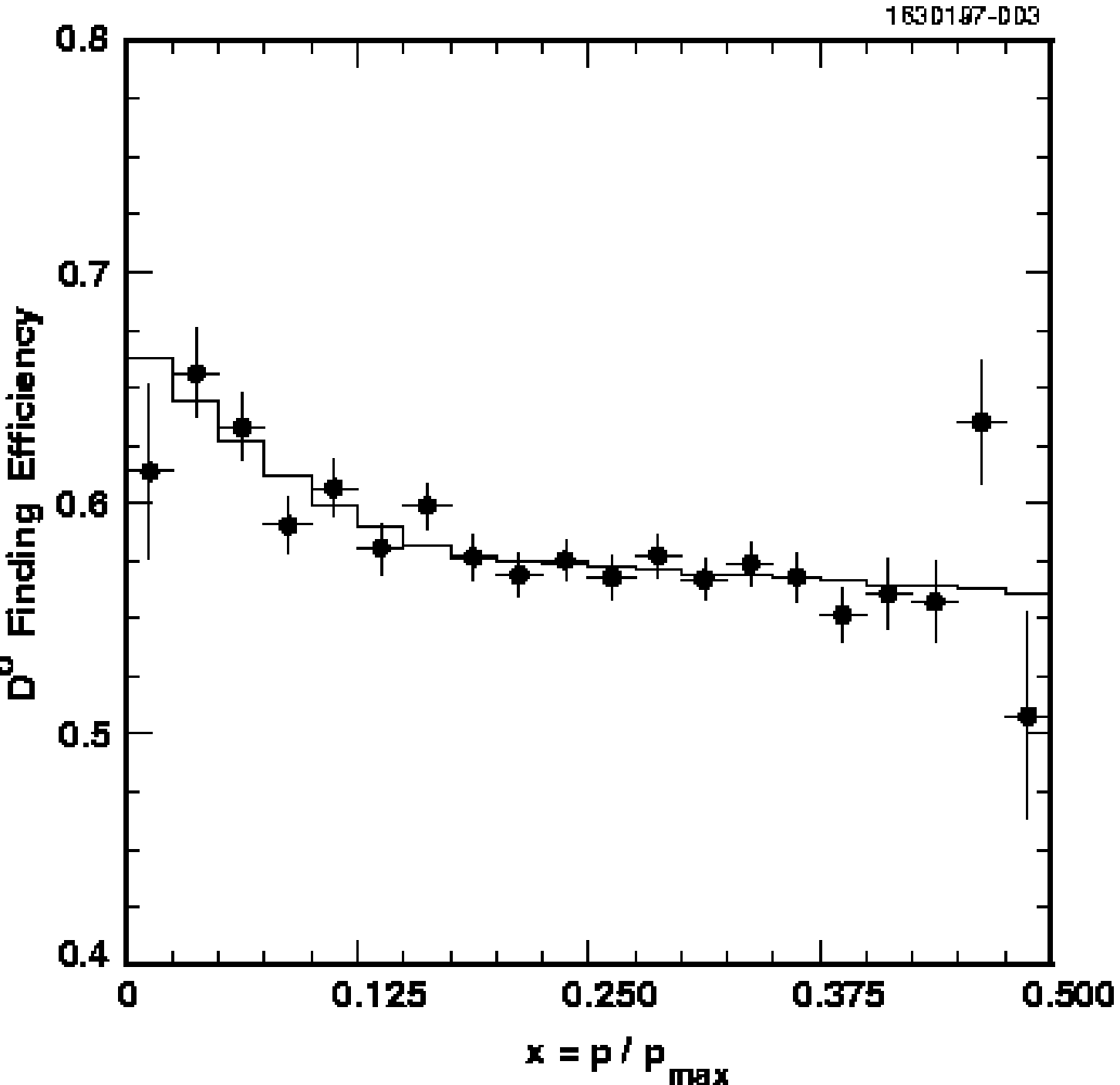,width=3.5in}}
\caption{\label{dz-eff} $D^0$ finding efficiency as a function of scaled
momentum $x$. The histogram is the result of the smoothing fit, binned.}
\end{figure}

The dedicated Monte Carlo simulated events were analyzed as real events to
produce the histogram of detection efficiency versus momentum that was
smoothed and used to correct the raw spectra (Fig.~\ref{dz-eff}).  The
detection efficiency is nearly independent of momentum in the region of
interest, except at the lowest momenta where there is an increase of angular
acceptance due to the near alignment of the two $D^0$ decay products.

\begin{figure}[htbp]
\centerline{\psfig{file=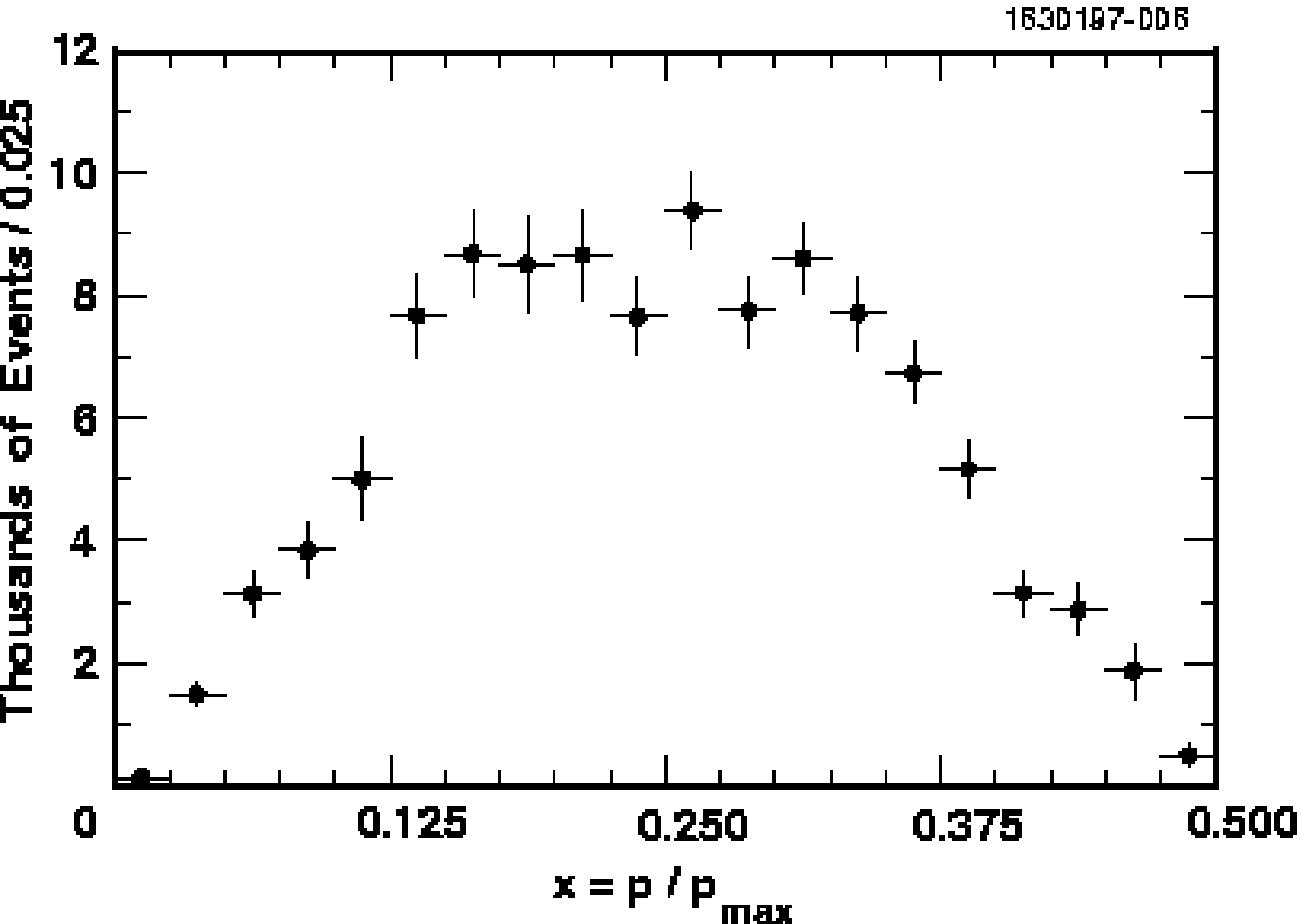,width=4.0in}}
\caption{\label{DZSPEC} The background subtracted and efficiency corrected
yield of $D^0$ mesons from $B\rightarrow D^0 X$ decay, as a function of
scaled momentum ($p_{max}=4.950~GeV/c$).} 
\end{figure}

The continuum-subtracted, efficiency-corrected, inclusive $D^0$ momentum
spectrum in $B\to D^0 X$ decay is shown in Fig.~\ref{DZSPEC}.  
Summing this spectrum over the interval $0<x<0.5$, we obtain the efficiency
corrected yield of $108,507\pm 2,407$ $D^0$s (resulting in a spectrum averaged
detection efficiency $= 0.578$) out of 4.3314 million $B$ decays.  This gives
the product of branching fractions,

\begin{equation}
{\cal B}_B{\cal B}_D  = (2.51\pm 0.06)\%, \label{dzprbr}
\end{equation}
where ${\cal B}_B\equiv{\cal B}(B\to D^0 X)$ and ${\cal B}_D\equiv{\cal
B}(D^0\to K^-\pi^+) + {\cal B}(D^0\to K^+\pi^-)$ \footnote{Our procedure finds
both the Cabibbo-allowed, $D^0\to K^-\pi^+$, and the doubly
Cabibbo-suppressed, $D^0\to K^+\pi^-$, decay modes, so we must divide by the
sum of the two decay branching fractions.} and the error is only statistical.

Using the CLEO results
${\cal B}(D^0\to K^-\pi^+) = (3.91\pm 0.08\pm 0.17)\%$ \cite{AKER} and
$\Gamma(D^0\to K^+\pi^-)/\Gamma(D^0\to K^-\pi^+) = (0.0077 \pm 0.0025\pm
0.0025)$ \cite{CINAB}, we obtain from Eq.~(\ref{dzprbr}) the branching
fraction 
\begin{equation}
{\cal B}(B\to D^0 X) = 0.636\pm 0.014 \pm 0.019 \pm 0.018. \label{BTODZ}
\end{equation}
The first error is statistical, while the second is the systematic error.
The contributions to the second (systematic) error, as percentages of the
branching fraction are given in table \ref{systmatic-d0}.
The third error is due to the error on 
the absolute branching fraction ${\cal B}(D^0\to K^-\pi^+)$. 

As part of a different project\cite{BKT}, an analysis of the $B\to D^0 X$
decay was carried out over a somewhat increased data sample and with more
stringent particle identification requirements, using also time of flight
information.  If both the $D^0\to K^-\pi^+$ and $\pi^- K^+$ hypotheses were
acceptable, but one gave a particle ID $\chi^2_{worse} > 4 + \chi^2_{better}$,
the hypothesis with the worse $\chi^2$ was rejected.  This selection in fact
reduced the switched mass background to zero.  The result of this analysis
for ${\cal B}(B\to D^0 X)$ is fully consistent with our result. 

\section{Inclusive $B\to D^+$ Decay}

\label{DPSEC} 

\begin{figure}[htbp]
\centerline{\psfig{file=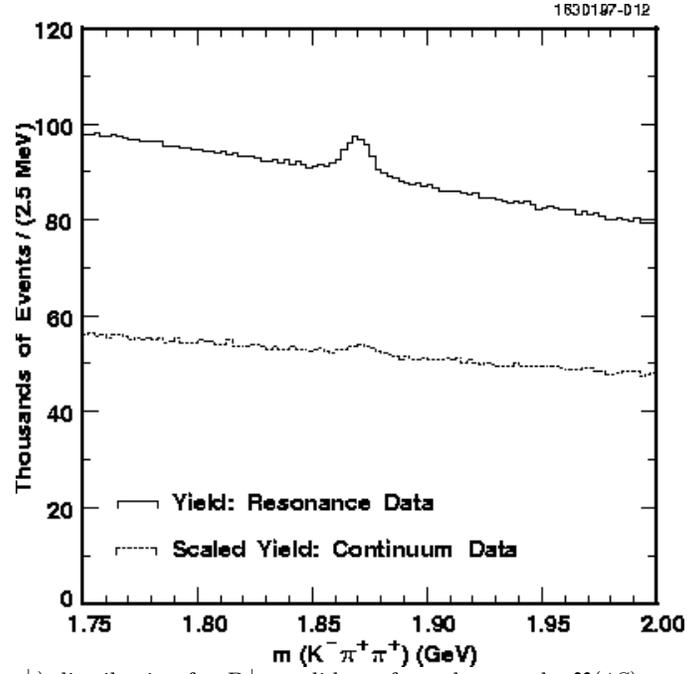,width=3.5in}}
\caption[dpglob]
{The $m(K^-\pi^+\pi^+)$ distribution for $D^+$ candidates from data at the
$\Upsilon(4S)$ resonance and from below $B\bar B$ threshold (scaled),
$0.0<x_D<0.5$.  The line is the result of the fit.}  
\label{dpglob}
\end{figure}

The complications from satellite peaks or switched mass backgrounds are not
present in the measurement of the $D^+$ spectrum and other kinematic
reflections are quite small (they will be discussed in the next subsection).  It
is then possible to estimate the shape of the combinatorial background by
fitting the $K^-\pi^+\pi^+$ mass distribution excluding the signal region.
However, it is difficult to determine the momentum dependence of the signal
shape parameters given the large combinatorial background and low signal to
background ratio.  The $K^-\pi^+\pi^+$ mass distributions summed over all
momenta of interest for the data at the $\Upsilon(4S)$ and those below $B\bar
B$ threshold (scaled according to luminosity and square of center-of-mass
energy ratios) are shown in Fig.~\ref{dpglob}.  If we parameterize the signal
as a double Gaussian, a modest change in the curvature of the combinatorial
background in the $m(K^-\pi^+\pi^+)$ distribution strongly correlates with a
wider and larger second component of the double Gaussian signal.

We analysed the $m(K^-\pi^+\pi^+)$ distributions using the double Gaussian
parameterization of the signal.  As in the analysis of the $D^0$
spectra, we used $D^+$ tagged by the $D^{*+}\to D^+\pi^0$ decay to 
determine the momentum dependence of the signal parameters.  We also
performed the analysis using the simple 
Gaussian signal parameterization.  This is quite adequate to fit
the data, given our statistical errors.  We found that the results from the
two different signal parameterizations are the same within a small fraction
of the statistical error.  Here we shall present the results obtained with
the simple Gaussian parameterization.

\subsection{Background from $D_s^+$ Decays}
We have small backgrounds from $D_s^+$ decays where a $K^+$ is misidentified
as a $\pi^+$.  We took this into account using the following procedure.  We
generated $B\bar{B}$ Monte Carlo events that contain at least one $D_s^+$,
which decays into $K^-\pi^+K^+$ directly or through the
$\overline{K^*}^0K^+$ or $\phi\pi^+$ resonance channel.  We processed them
through the simulation of our detector and track reconstruction.  We selected
events that passed the selection criteria for the decay $D^+\to
K^-\pi^+\pi^+$.  We plotted these (fake) $m(K^-\pi^+\pi^+)$ distributions for
each momentum bin and normalized them using our knowledge of the differential
${\cal B}(B\to D_s^+ X)$\cite{DSMEN} and of the $D_s$ decay branching ratios.
We added these histograms to the combinatorial background and signal
functions when fitting the data.  Taking into account this feed-down we changed
our result for ${\cal B}(B \to D^+ X)$ by $\Delta{\cal B} = -0.0021\pm
0.0002$.  We have examined other possible feed-downs and verified that
they do not affect our measurement.

\subsection{$D^+$ spectrum}

\begin{figure}[htbp]
\centerline{\psfig{file=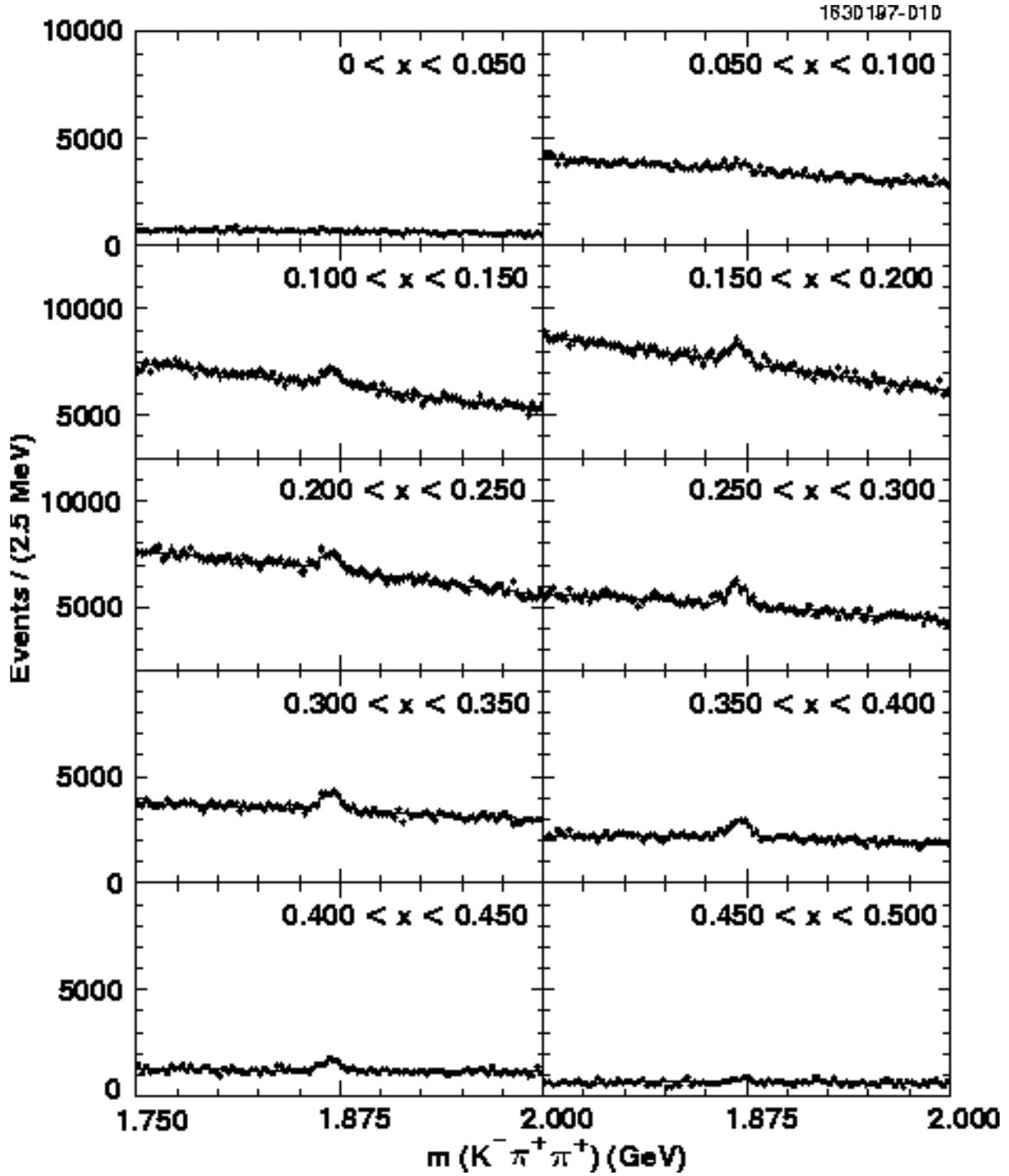,width=6.0in}}
\caption{The $m(K^-\pi^+\pi^+)$ distribution for $B\to D^+\ X$ candidates in ten
 $x$ bins from 0.0 to 0.50.\label{figDplus}} 
\end{figure}

\begin{figure}[htbp]
\centerline{\psfig{file=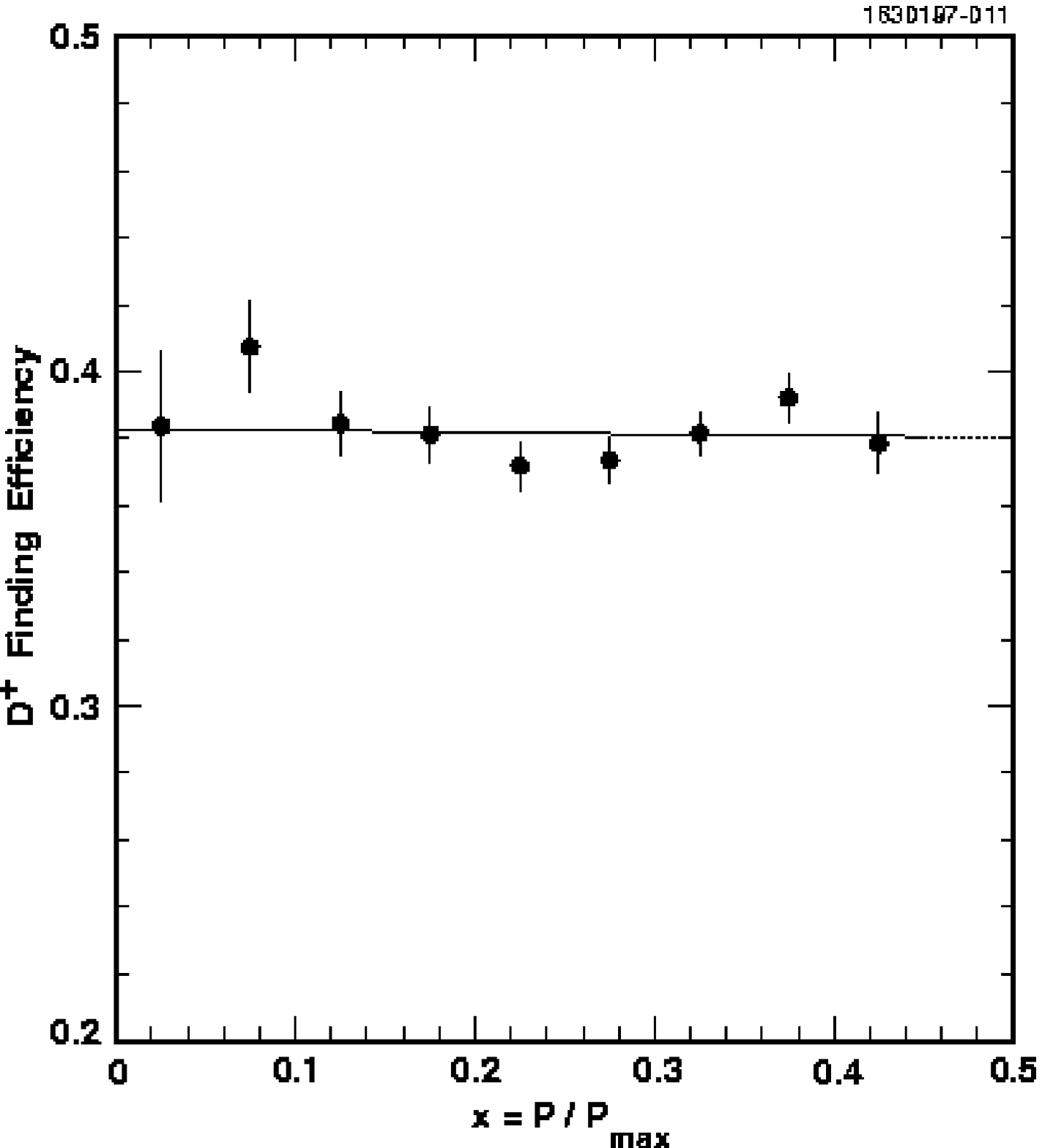,width=3.5in}}
\caption[DP-EFF]{\label{dp-eff} $D^+$ finding efficiency as a function 
of scaled momentum $x$. The histogram is the result of the smoothing fit,
binned.} 
\end{figure}

We fitted the continuum subtracted sample of $D^+\to K^-\pi^+\pi^+$
candidates.  We divided 
the sample in 20 momentum bins between $0.00<x<1.00$ and performed the
bin-by-bin fit of the $m(K^-\pi^+\pi^+)$ distributions using the signal and
background functions described above.
In Fig.~\ref{figDplus} we show the fitted mass distributions in the 10 $x$
bins.  The sum over the 10 bins for $0.0<x_D<0.5$ gave a raw yield of
$35,804\pm 1,297$ $D^+$ from $B$ decay.  The detection efficiency as a
function of $x$ is shown in Fig.~\ref{dp-eff}.  We have smoothed the $x$
dependence in different ways.  The resulting average efficiencies differ from
one another by 0.18\% at most.  The error introduced by the choice of the
smoothing function has been taken into account in the systematic error.  The
smoothing actually used is shown in Fig.~\ref{dp-eff}.

\subsection{Result}
\begin{figure}[htbp]
\centerline{\psfig{file=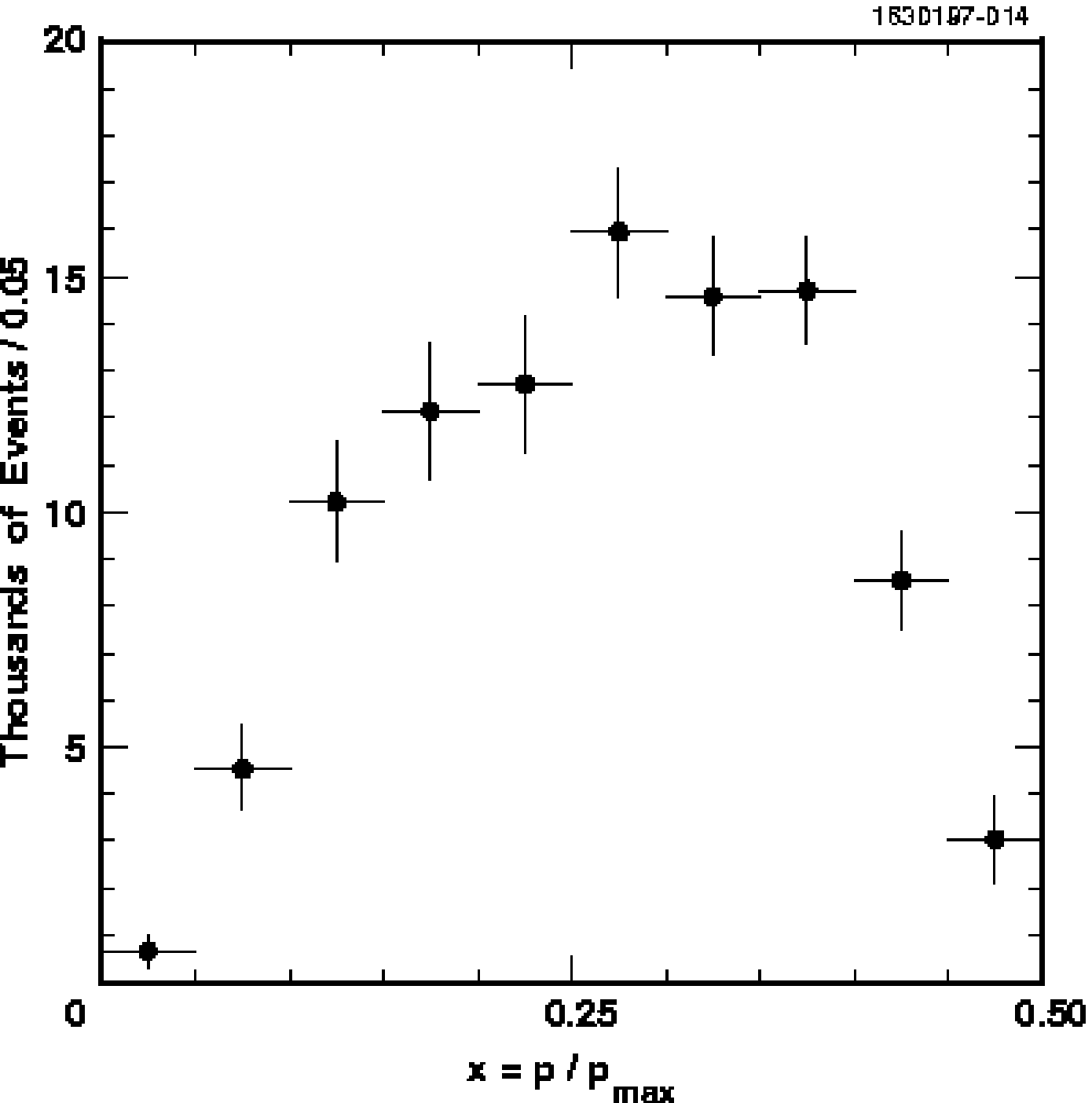,width=3.5in}}
\caption{The background subtracted and efficiency corrected
yield of $D^+$ mesons from $B\to D^+ X$ decay, as a function of
scaled momentum ($p_{max}=4.950~GeV/c$).} 
\label{dpspec} 
\end{figure}

The efficiency corrected $D^+$ spectrum is shown in Fig.~\ref{dpspec}.
Summing this spectrum over the interval $0.0<x_D<0.5$, we obtained an
efficiency corrected yield of $93,493\pm 3,386$ $D^+$ (resulting in a
spectrum averaged detection efficiency $= 0.383$) that gives the product of
branching fractions,

\begin{equation}
{\cal B}_B{\cal B}_D = (2.16\pm 0.08)\%, \label{dpprbr}
\end{equation}

\noindent where ${\cal B}_B\equiv{\cal B}(B\to D^+ X)$, ${\cal
B}_D\equiv{\cal B}(D^+\to K^-\pi^+\pi^+)$ and the error is statistical only.

Using CLEO's result for the absolute branching fraction
${\cal B}(D^+\to K^-\pi^+\pi^+) = (9.19\pm0.6\pm0.8)\%$ \cite{BALE} gives:

\begin{equation} 
{\cal B}(B \to D^+ X) = 0.235\pm 0.009\pm 0.009 \pm 0.024. \label{BTODP}
\end{equation}

The first error is statistical, the second is systematic, and the third error
is due to the statistical error on ${\cal B}(D^+\to K^-\pi^+\pi^+)$ and to
that part of its systematic error that propagates to our measurement.
Table~\ref{systmatic-dp} gives the components of the second error as
percentages of the branching fraction.

Also this analysis, as in the $B\to D^0\ X$ case, has been carried out with
different procedures and selection criteria that produced results within a
fraction of the systematic error of the one reported here.  

We will show later that the $B\to D^{*+} X$ branching fractions measured by
reconstructing $D^{*+}$ through the decay chains $D^{*+}\to D^0\pi^+\to
(K^-\pi^+)\pi^+$ and $D^{*+}\to D^+\pi^0\to (K^-\pi^+\pi^+)\pi^0$ are
consistent with each other, their ratio being $1.01\pm 0.09$. This is a
consistency check of the results given above.

\section{Inclusive $B\to D^*$ Decay}
\label{DSSEC} 
We selected $D^*$ candidates by combining a $\pi$ of the appropriate charge
with a $D$ candidate and then imposing requirements on $\delta m$, the
difference between the mass of the $D^*$ candidate and that of the $D$
candidate.  However, this sample also contains background composed of pions
randomly associated with a correctly reconstructed $D$ meson.  We eliminated
this ``fake'' $D^*$ 
background by subtracting properly scaled $\delta m$\ ``sidebands'' as
described in Sec.~\ref{DSTAG}.  The mass distribution of the $D$ candidates
so obtained was fitted to find the $D$ yield as in the previous sections.
\subsection{Inclusive $B\to D^{*0}$ Decay}
\begin{figure}[htbp]
\centerline{\psfig{file=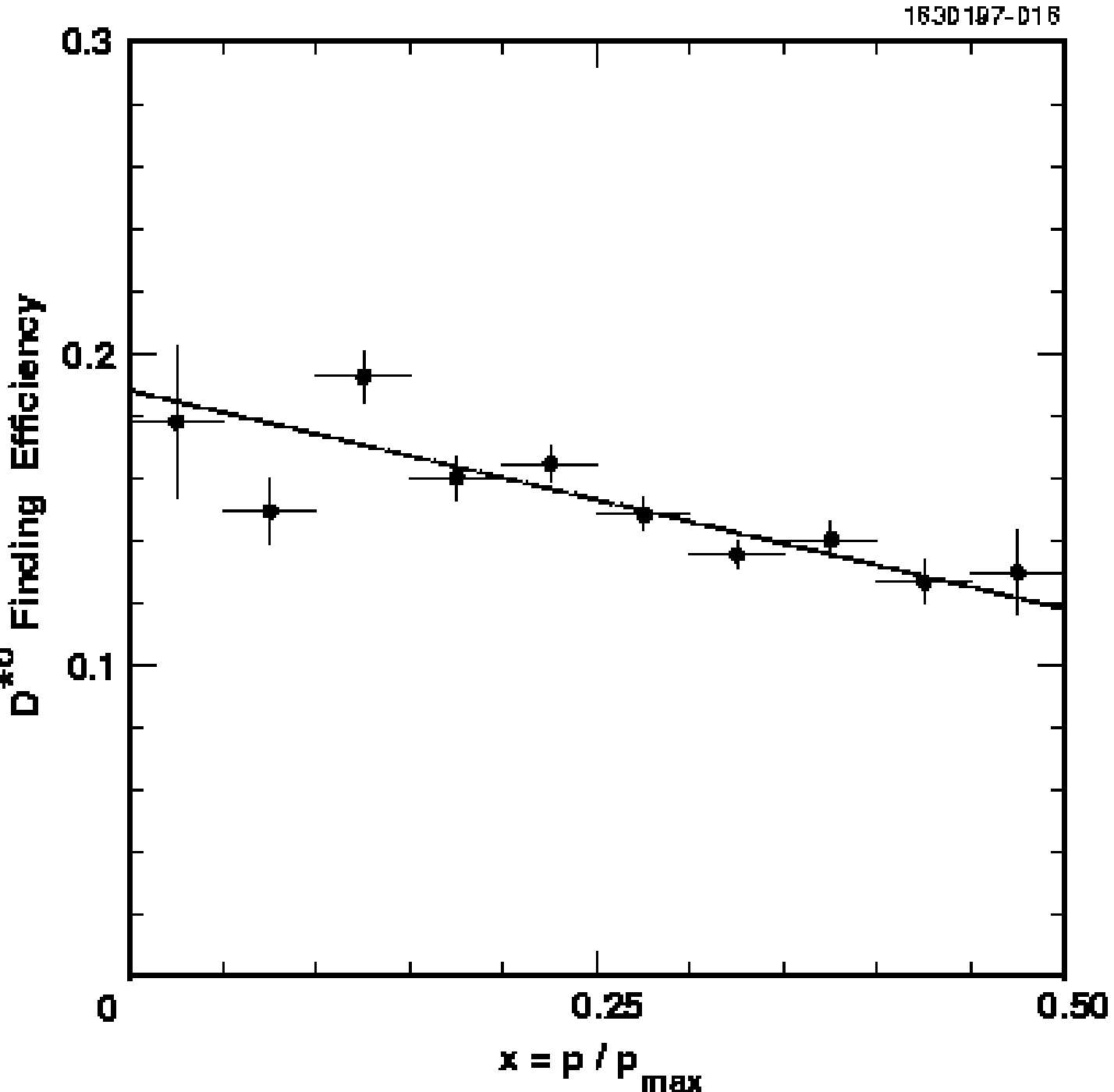,width=3.5in}}
\caption{$D^{*0}$ finding efficiency as a function of scaled momentum $x$. 
The histogram is the result of the smoothing fit, binned.}
\label{DSZEFF}
\end{figure}

\begin{figure}[htbp]
\centerline{\psfig{file=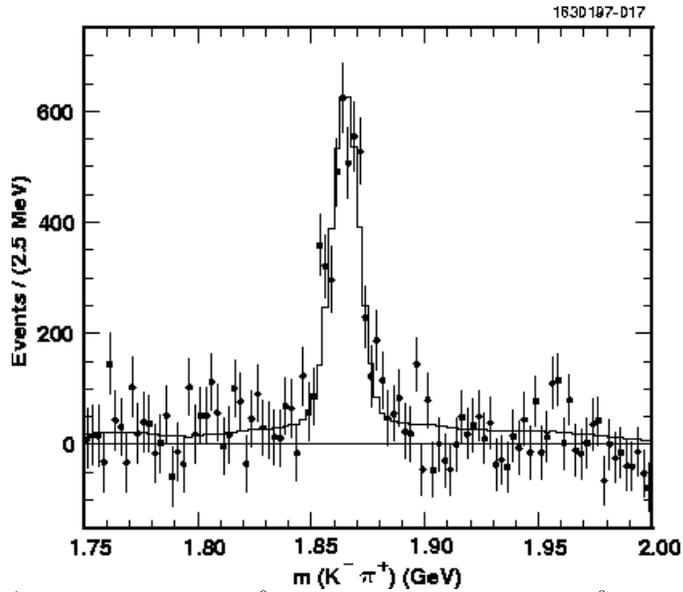,width=3.5in}}
\caption{The $m(K^-\pi^+)$ distribution for $D^0$ candidates from $B\to D^{*0}\
 X$ in the momentum interval $(0.0<x_{D^*}<0.50).$
\label{dsz2}}
\end{figure}

\begin{figure}[htbp]
\centerline{\psfig{file=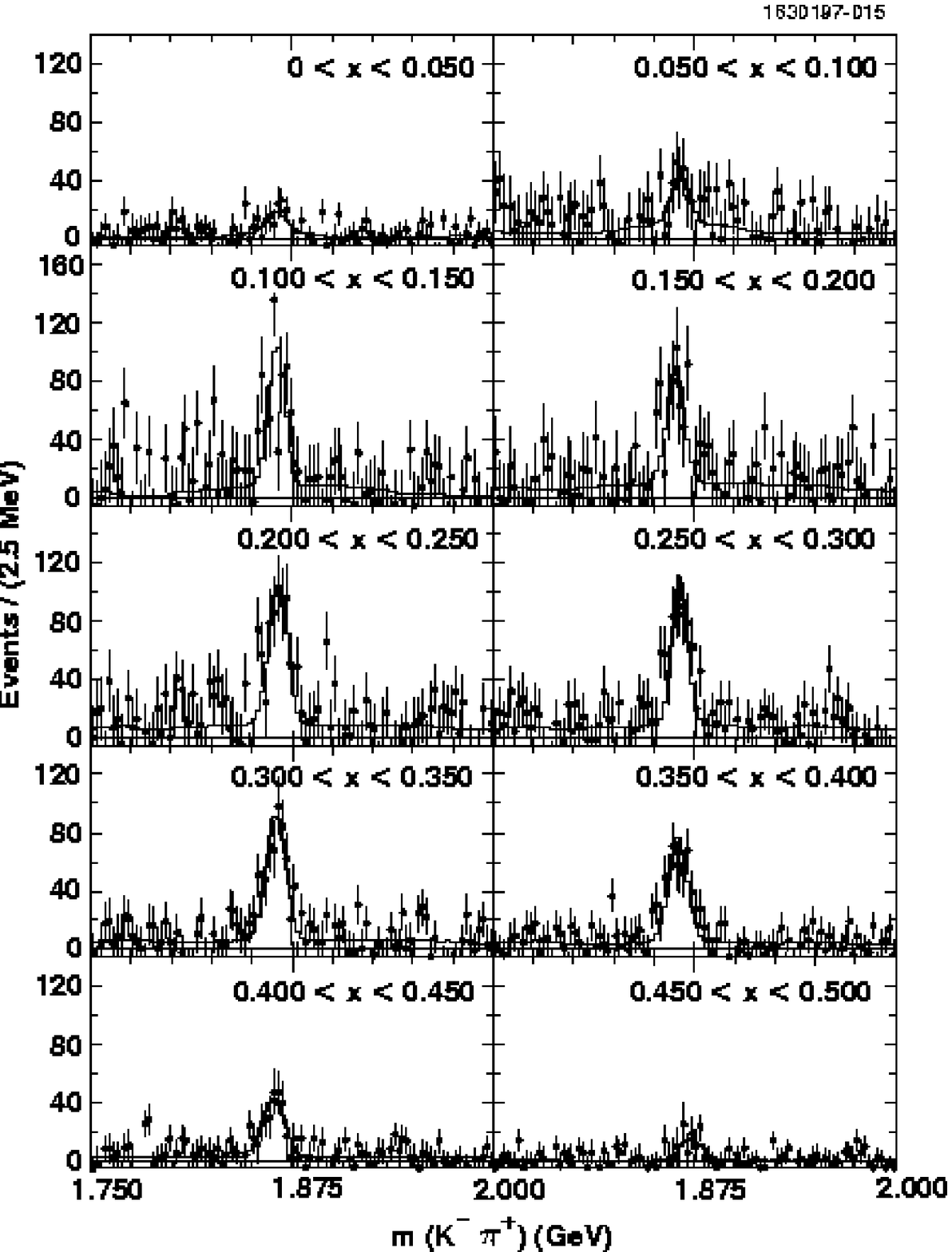,width=6.0in}}
\caption{The $m(K^-\pi^+)$ distribution for $D^{*0}$ in ten $x$ bins from 0 to
 0.50.  The line is the result of the fit.}  
\label{figDstar0}
\end{figure}

\begin{figure}[htbp]
\centerline{\psfig{file=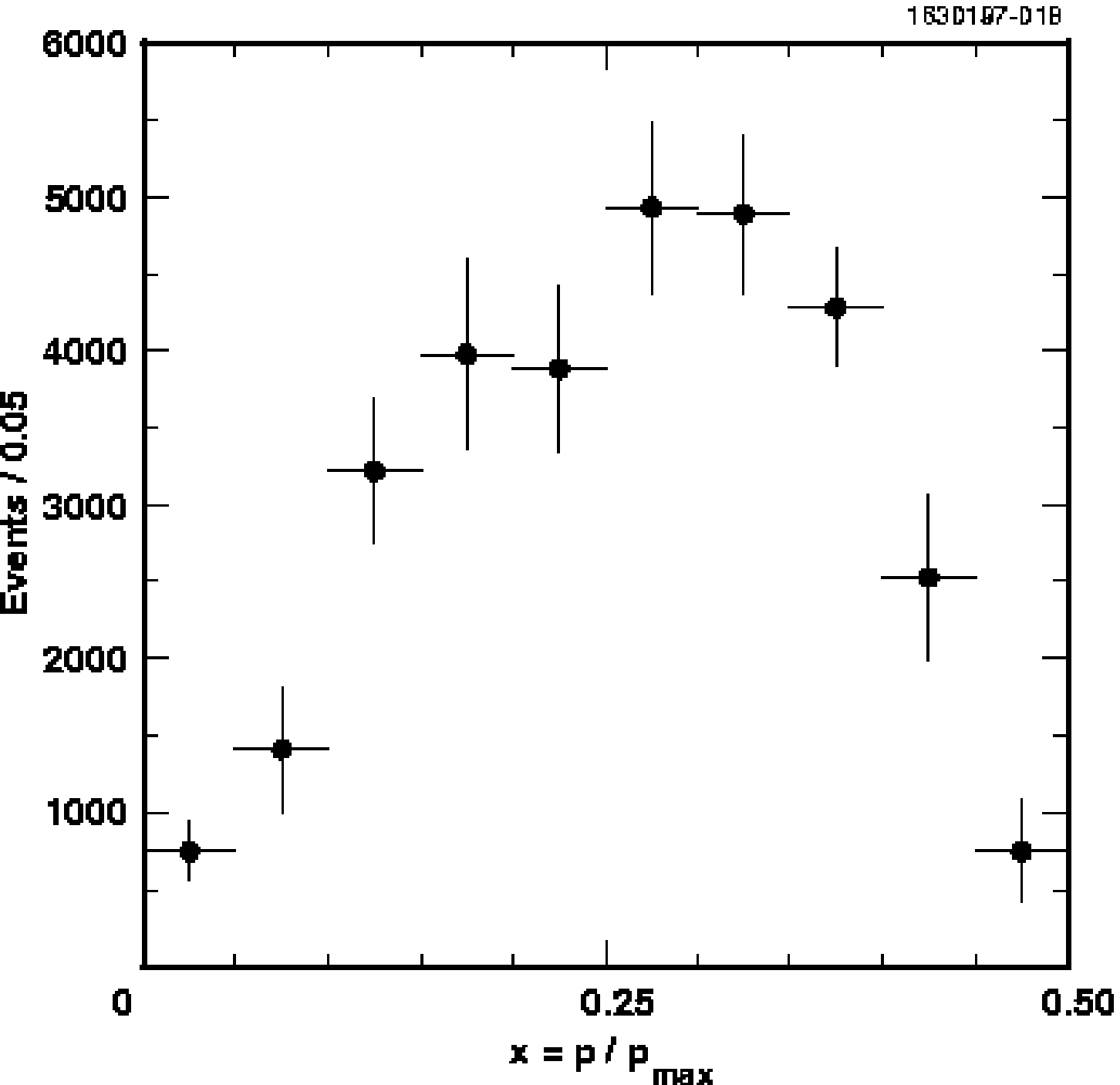,width=3.5in}}
\caption{The background subtracted and efficiency corrected yield of $D^{*0}$
mesons from $B\rightarrow D^{*0} X\rightarrow (D^0\pi^0) + X$ decay, as a
function of scaled momentum ($p_{max}=4.950~GeV/c$).} 
\label{dsz1}
\end{figure}

We selected $D^0$ candidates in the decay mode $D^0 \to K^- \pi^+$ (as
described in Sect.~\ref{DZSEC}) and combined them with a $\pi^0$ to form
$D^{*0}$ candidates.  The $D^{*0}$ yield in each momentum bin was found by
fitting the $m(K^-\pi^+)$ distribution obtained with the $\delta m$ selection
and sideband subtraction procedure.  This decay channel is affected by the
same backgrounds as the $D^0$ channel.  To fit the $m(K^-\pi^+)$ mass
distributions, we used the parameterization and procedures from the analysis
of $B\to D^0~+~X$ described in Sect.~\ref{DZSEC}.  However, the large number
of free parameters involved with the multi-Gaussian parameterization was not
suitable with the lower statistics available in this channel.  We have thus
used the histograms of the satellite peaks and of the switched $K-\pi$
distributions from Monte Carlo tagging while for the signal we used either a
single Gaussian or, the histogram from Monte Carlo tagging.  
Both procedures
gave ${\cal B}(B\to D^{*0}~+~X)$ values differing by a small fraction of the
statistical error.  We report here the results from the single Gaussian
parameterization of the signal.  In Fig.~\ref{figDstar0} we show the fitted
mass distributions in the 10 $x$ bins.  The sum over the 10 bins gave a raw
yield of $3,539\pm 175$ $D^{*0}$ from $B$ decay.  After correcting bin-by-bin
for detection efficiency (Fig.~\ref{DSZEFF}), we obtained the inclusive
$D^{*0}$ spectrum shown in Fig.~\ref{dsz1}.  Fig.~\ref{dsz2} shows the
continuum subtracted $K^-\pi^+$ effective mass distribution of $D^{*0}$
candidates from the decay chain $B\to D^{*0} X \to (D^0 \pi^0) X$ in the
momentum interval $(0.0<x_{D^*}<0.50)$.

As in the previous cases, by summing the spectrum in the interval $0.0<x<0.5$
we found a corrected yield of $26,840\pm 1,331$ $D^{*0}$ from $B$ decay
(resulting in a spectrum averaged detection efficiency $= 0.132$) and the
product of branching fractions,

\begin{equation}
{\cal B}_B{\cal B}_D^*{\cal B}_D = (0.620\pm 0.031)\% \label{dszbbb}
\end{equation}

\noindent where ${\cal B}_B\equiv{\cal B}( B \to D^{*0} X)$,
${\cal{B}}_D^*\equiv{\cal B}(D^{*0}\to D^0\pi^0)$,
${\cal{B}}_D\equiv{\cal{B}}(D^0\to K^-\pi^+)$  and only the statistical error
is shown. 

Using the CLEO results
${\cal B}(D^{*0}\to D^0\pi^0) = (63.6\pm 2.3\pm 3.3)\%\cite{BUT}$ and
${\cal B}(D^0\to K^-\pi^+)\ = (3.91\pm 0.08\pm 0.17)\%$ \cite{AKER}, and
$\Gamma(D^0\to K^+\pi^-)/\Gamma(D^0\to K^-\pi^+) = (0.0077 \pm 0.0025\pm
0.0025)$ \cite{CINAB},     
from Eq.~(\ref{dszbbb}) we obtain the branching fraction:

\begin{equation}
{\cal B}(B\to D^{*0} X) = 0.247\pm 0.012\pm 0.018\pm 0.018. \label{BTODSZ} 
\end{equation}

\noindent The first error is statistical. The second error is systematic and
includes the components listed in Table~\ref{systmatic-f0}.  These are given
as percentages of the branching fraction itself.  The third error is
determined from the statistical error on ${\cal B}(D^0\to K^-\pi^+)$.

\subsection{Inclusive $B\to D^{*+}$ Decay}
\label{DSPSEC} 
\begin{figure}[htbp]
\centerline{\psfig{file=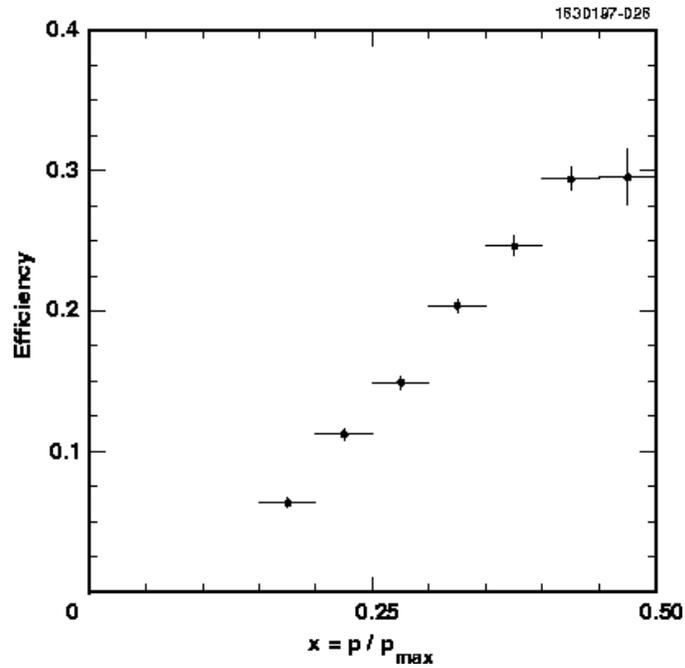,width=3.5in}}
\caption{The $D^{*+} \rightarrow D^{0} \pi^{+}$ finding efficiency as a
function of scaled momentum $x$.} 
\label{DSPEFFA}
\end{figure}

\begin{figure}[htbp]
\centerline{\psfig{file=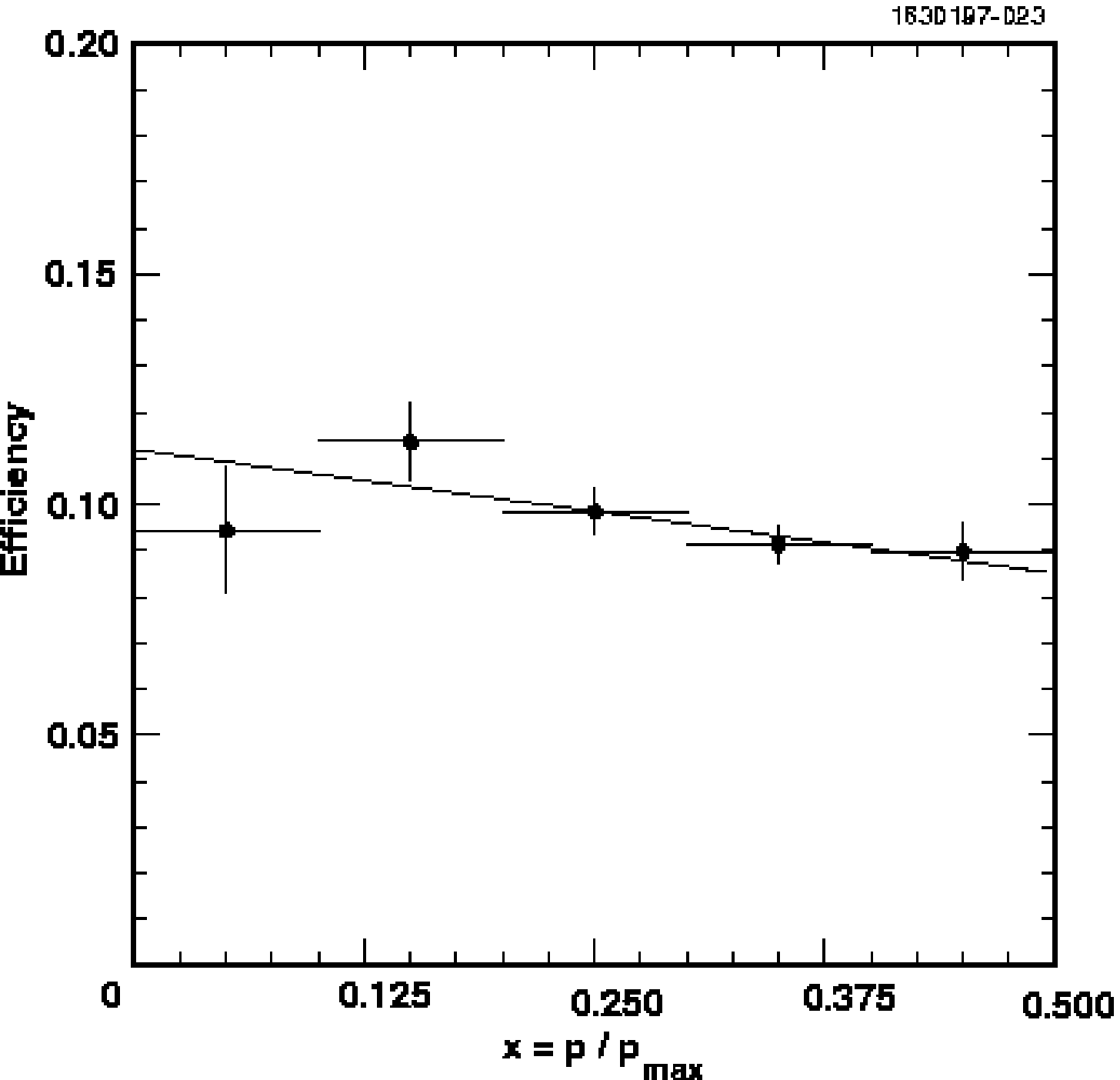,width=3.5in}}
\caption{$D^{*+}\rightarrow D^+\pi^0$ finding efficiency as a function of the
scaled momentum $x$. The histogram is the result of the smoothing fit,
binned.}  
\label{DSPEFFB}
\end{figure}

We have analysed the inclusive decay $B\to D^{*+}$ in two $D^{*+}$ decay
modes.  The $D^{*+}\to D^0\pi^+$ channel has significantly more events
because of the larger $D^*$ branching fraction and higher detection
efficiency.  However, the detection efficiency is a steep function of the
$D^{*+}$ momentum (Fig.~\ref{DSPEFFA}) because of the short range of the low
momentum $\pi^+$ and absorption in the beampipe.  The detection efficiency
for the charged pion is nearly zero for $x_{D^*}<0.15$.  Summing over the
spectrum gives ${\cal B}( B\to D^{*+} X)$ in the charged pion mode for
$x_{D^*}>0.15$.  The $D^{*+}\to D^+\pi^0$ channel, which has fewer events and
much larger backgrounds (Fig.~\ref{DELMFPC}) has an efficiency which is
nearly constant with momentum (Fig.~\ref{DSPEFFB}) and is the only source of
information for the low momentum region.  We will separately describe the
analyses using the two $D^{*+}$ decay modes and then discuss how to combine
them to obtain ${\cal B}(B\to D^{*+} X)$.

\subsubsection{Using the $D^{*+}\to D^0\pi^+$ decay channel}

\label{DSPSECA} 
\begin{figure}[htbp]
\centerline{\psfig{file=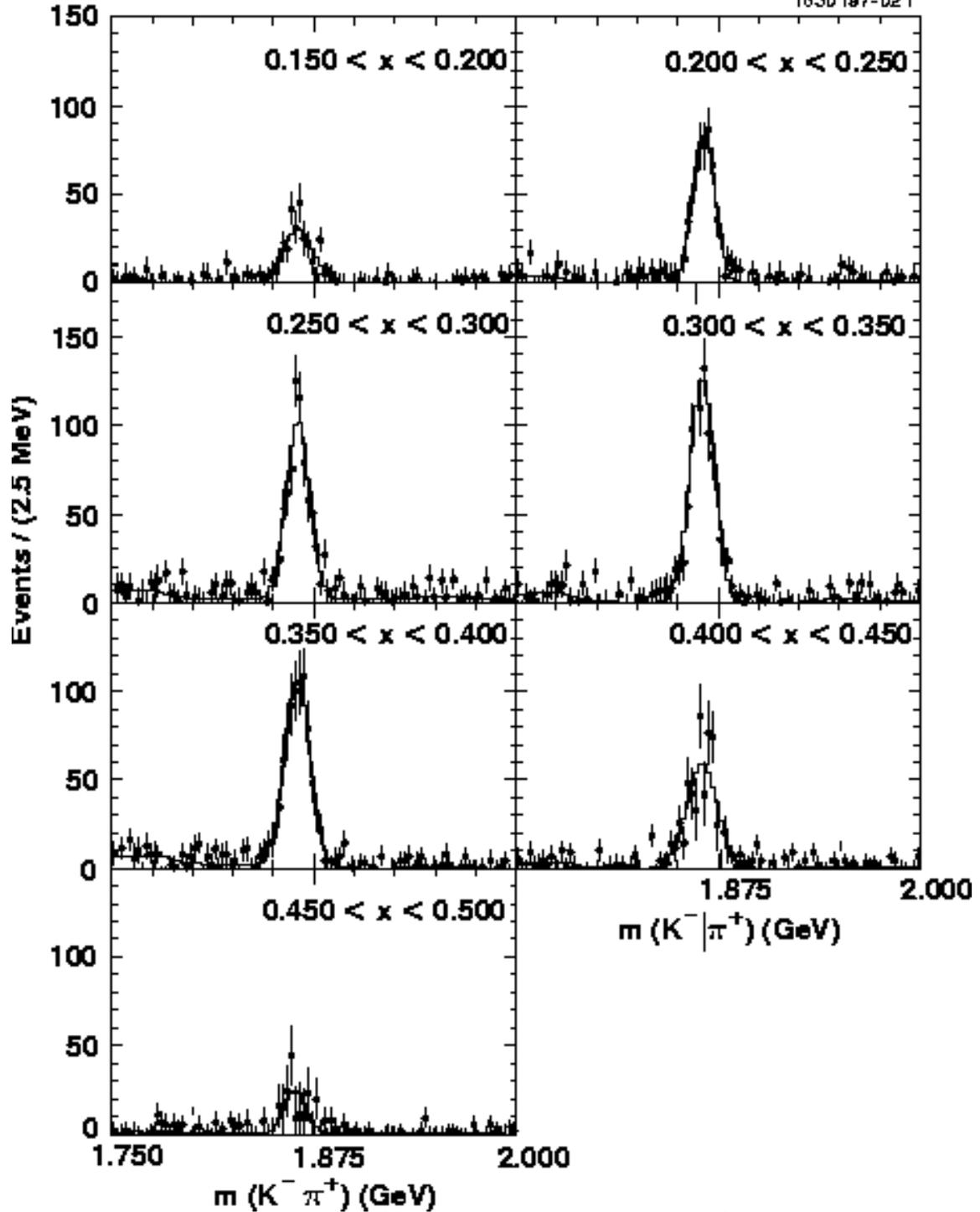,width=6.0in}}
\caption{The $m(K^-\pi^+)$ distribution for $D^0$ candidates from $B\to D^{*+}\
 X \to (D^0\pi^+)\ X$ in 7 $x$ bins from 0.15 to 0.50.  The $\delta m$
 sidebads have been subtracted. The line is the result of the fit.}
\label{figDstarPlus}
\end{figure}

Candidate $D^0$ mesons were reconstructed in the $D^0\to K^-\pi^+$ decay mode
(as described in Sec.~\ref{DZSEC}) and combined with a $\pi^+$ (referred to
as the ``slow $\pi$'') to form $D^{*+}$ candidates.  In order to maximize
the detection efficiency of the slow $\pi^+$ we did not require $dE/dx$
information to be available for this track.  The $D^{*+}$ yield in each of
the 7 momentum bins $(0.15<x_{D^*}<0.50)$ was found by fitting the
$m(K^-\pi^+)$ distribution obtained through the $\delta m$\ selection and
sideband subtraction procedure described in Sec.~\ref{DSTAG}.  In fitting
the $D^0$\ peak, we used the same parameterization that was used in the
analysis of $B\to D^0~+~X$, but no background due to double misidentification
of $D^0\to K^-\pi^+$ is present in this mode.  In Fig.~\ref{figDstarPlus} we
show the fitted mass distributions in the 7 $x$ bins.  Summing the spectrum
in the interval $0.15<x_{D^*}<0.5$, gave a raw yield of $4,214\pm 120$
$D^{*+}$.  

\begin{figure}[htbp]
\centerline{\psfig{file=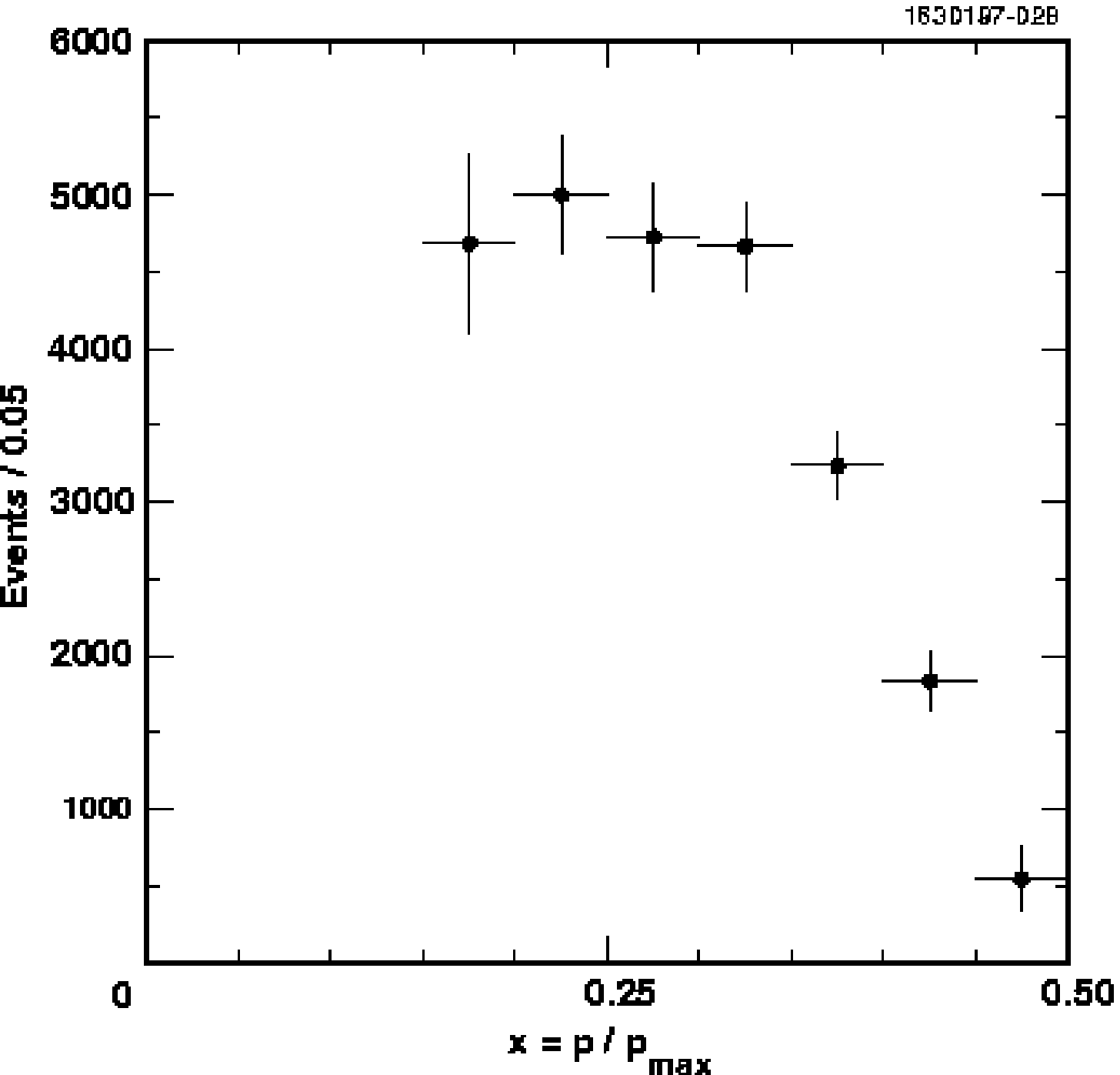,width=3.5in}}
\caption{The background subtracted and efficiency corrected yield of $D^{*+}$
mesons from $B\to D^{*+} X\to (D^0\pi^+)\ X$ decay, as a
function of scaled momentum  ($p_{max}=4.950~GeV/c$).} 
\label{dspp1}
\end{figure}

Correcting bin-by-bin for detection efficiency (Fig.~\ref{DSPEFFA}), we
obtained the inclusive $D^{*+}$ spectrum from $B$ decay shown in
Fig.~\ref{dspp1}.  The efficiency corrected yield for $x_{D^*}>0.15$ is
$24,691\pm 902$ $D^{*+}$ from $B$ decay (resulting in a spectrum averaged
detection efficiency $= 0.171$). This gives the product of branching
fractions,

\begin{equation}
{\cal B}_B{\cal B}_D^*{\cal B}_D(x_{D^*}>0.15) = (0.570\pm 0.021)\%
\end{equation}

\noindent where ${\cal B}_B\equiv{\cal B}( B\to D^{*+} X)$,
${\cal{B}}_D^*\equiv{\cal{B}}(D^{*+}\to D^0\pi^+)$, 
${\cal{B}}_D\equiv({\cal B}(D^0\to K^-\pi^+)+{\cal B}(D^0\to K^+\pi^-))$ and
the error quoted is only statistical. 

\begin{figure}[htbp]
\centerline{\psfig{file=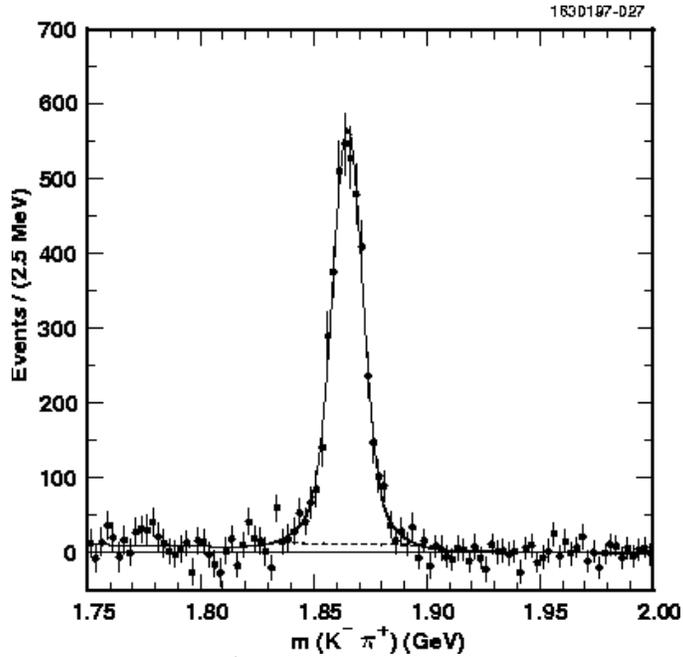,width=3.5in}}
\caption{The $m(K^-\pi^+)$ distribution for $D^{*+}$ candidates from $B$ decay
$(0.15<x_{D^*}<0.50)$, after $\delta m$\ sideband subtraction.  The dashed
line shows the background under the signal.}
\label{dspp2}
\end{figure}

Fig.~\ref{dspp2} shows the continuum subtracted $K^-\pi^+$ effective mass
distribution of $D^{*+}$ candidates from the decay chain 
$B\to D^{*+}\ X\to (D^0\pi^+) X$ 
in the momentum interval interval $(0.15<x_{D^*}<0.50)$, after $\delta m$\
sideband subtraction.

\subsubsection{Using the $D^{*+}\to D^+\pi^0$ decay channel}

\label{DSPSECB} 
\begin{figure}[htbp]
\centerline{\psfig{file=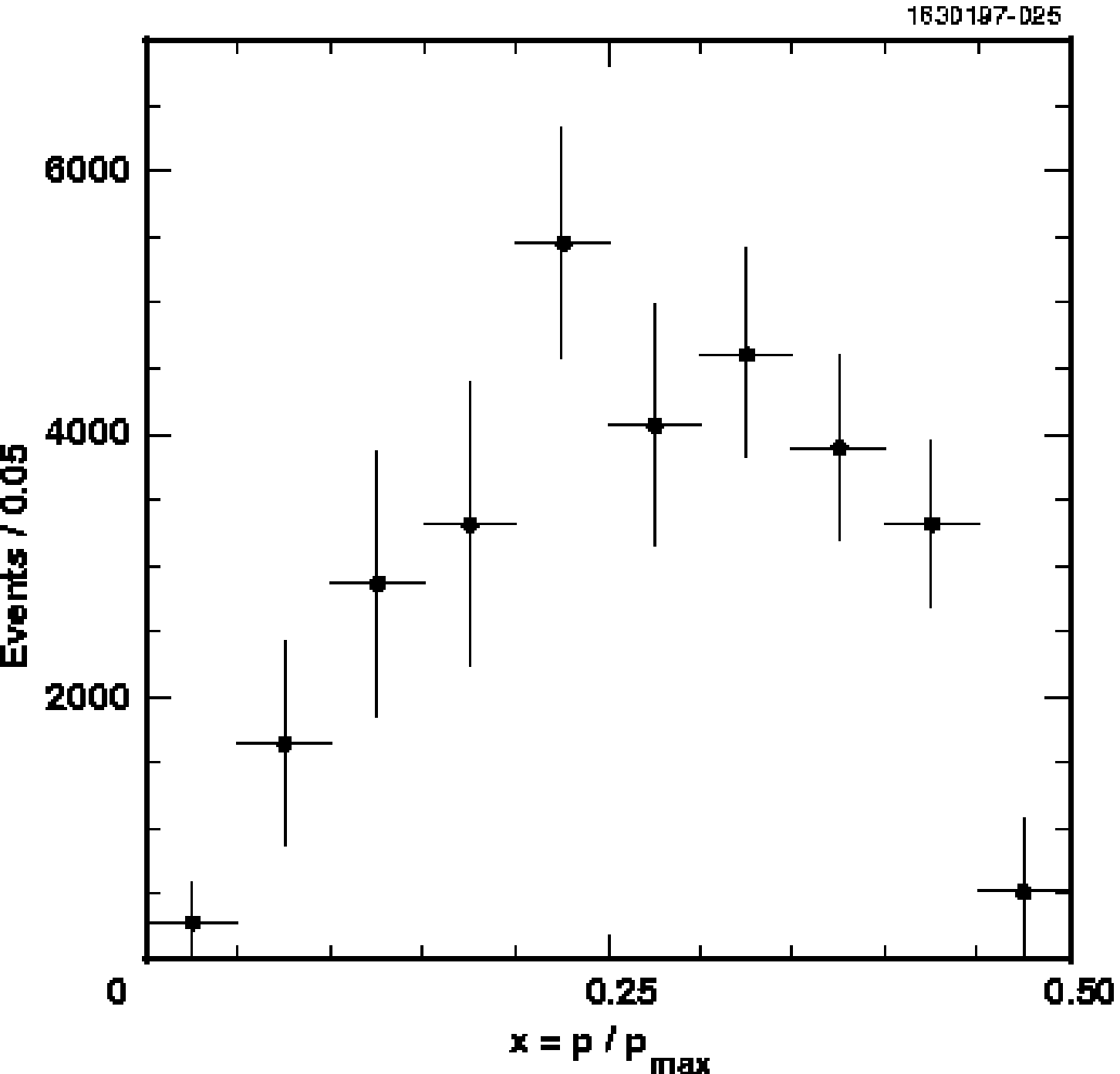,width=3.5in}}
\caption{The background subtracted and efficiency corrected
yield of $D^{*+}$ mesons from $B\rightarrow D^{*+} X\rightarrow (D^+\pi^0)\ X$
decay, as a function of scaled momentum ($p_{max}=4.950~GeV/c$).}
\label{DSPZ1} 
\end{figure}

Candidate $D^+$s are reconstructed in the decay mode $D^+\to K^-\pi^+\pi^+$
(as described in Sec.~\ref{DPSEC}) and combined with a $\pi^0$ to form
$D^{*+}$ candidates.  The $D^{*+}$ yield in each of the 20 momentum bins
$(0.0<x_{D^*}<1.0)$ was found by fitting the $m(K^-\pi^+\pi^+)$ distribution
obtained from the $\delta m$ selection and sideband subtraction procedure
described in Sec.~\ref{DSTAG}.  The $D^+$ peak was fitted using the same
parameterization and procedure used in the analysis of $B\to D^+ X$.  Summing
the spectrum over the interval $0.0<x_{D^*}<0.5$ gave a raw yeld of $2,925\pm
250$ $D^{*+}$.  After correcting for detection efficiency
(Fig.~\ref{DSPEFFB}) and summing over x bins, we obtained the inclusive
$D^{*+}$ spectrum from $B$ decay shown in Fig.~\ref{DSPZ1}.  The corrected
yield is $27,683\pm 2,339$ $D^{*+}$ from $B$ decay (resulting in a spectrum
averaged detection efficiency $= 0.106$) and the product of branching
fractions is,

\begin{equation}
{\cal B}_B{\cal B}_D^*{\cal B}_D = (0.639\pm 0.054)\%
\end{equation}

\noindent where ${\cal B}_B\equiv{\cal B}(B\to D^{*+} X)$,
${\cal B}_D^*\equiv{\cal{B}}(D^{*+}\to D^+\pi^0)$,
${\cal B}_D\equiv{\cal B}(D^+\to K^-\pi^+\pi^+)$ and the error quoted is
statistical only. 

\begin{figure}[htbp]
\centerline{\psfig{file=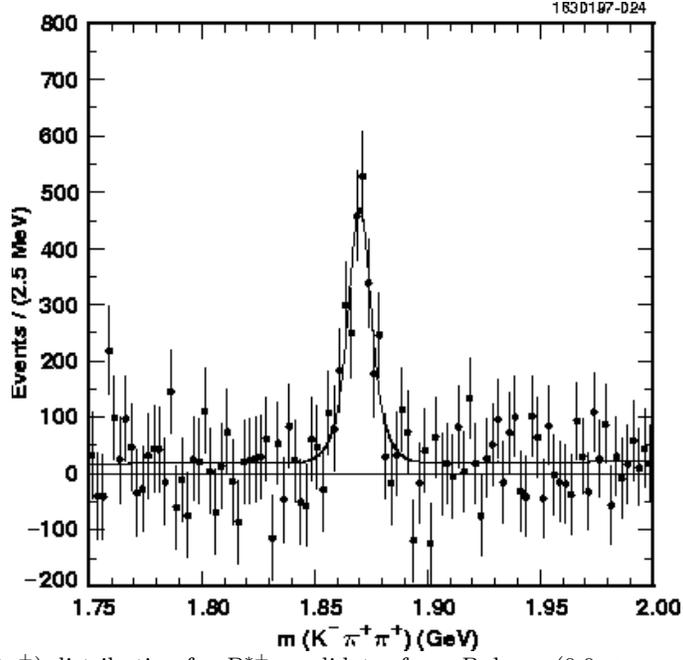,width=3.5in}}
\caption{The $m(K^-\pi^+\pi^+)$ distribution for 
$D^{*+}$ candidates from $B$ decay $(0.0<x_{D^*}<0.50)$, after $\delta m$\
sideband subtraction.}
\label{DSPZ2}\end{figure}

Fig.~\ref{DSPZ2} shows the continuum subtracted $K^-\pi^+\pi^+$ invariant
mass distribution of $D^{*+}$ candidates from the decay chain $B\to D^{*+} +
X\to (D^+\pi^0) X$ in the whole momentum interval interval
$(0.0<x_{D^*}<0.50)$.

\subsection{Combined results for the inclusive $B\to D^{*+}$ Decay}

\begin{figure}[htbp]
\centerline{\psfig{file=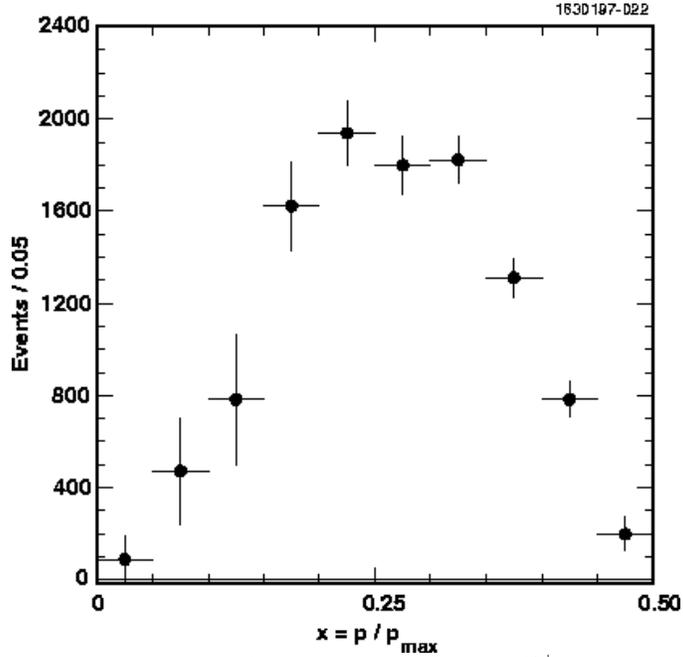,width=3.5in}}
\caption{The background subtracted and efficiency corrected yield of $D^{*+}$
mesons from $B\rightarrow D^{*+} X$ decay, as a function of scaled momentum
($p_{max}=4.950~GeV/c$).  For $0<x<0.15$, only the $D^{*+} \rightarrow D^{+}
\pi^{0}$ measurement is used.  For $0.15<x<0.5$, measurements from both
$D^{*+} \rightarrow D^{0}\pi^{+}$ and $D^{*+} \rightarrow D^{+} \pi^{0}$ are
combined.} 
\label{dsspec2}
\end{figure}

Using the corrected differential branching fractions $d{\cal B}(B\rightarrow
D^{*+} X)/dx$ obtained in two independent $D^*$ decay modes, we combined them
as follows.  In the momentum region $0.0<x_{D^*}<0.15$ we use the only
measurement available, that from the $D^{*+}\to D^+\pi^0$ decay mode, ${\cal
B}(B\rightarrow D^{*+} X)(0.0<x_{D^*}<0.15) = 0.031\pm 0.009\pm 0.0025\pm
0.0027$.  In the momentum region $0.15<x_{D^*}<0.50$ we calculated the
weighted average of the two measurements.  The resulting spectrum is shown in
Fig.~\ref{dsspec2}.

The sum over all $x$ bins gives the branching fraction, 
\begin{equation}
{\cal B}(B \to D^{*+} X) = 0.239\pm 0.011\pm 0.014\pm0.009. \label{BTODSP} 
\end{equation}
The first error is statistical and is dominated by the error on the 
$D^*$ branching fraction 
for $0.0<x_{D^*}<0.15$.  The second error is systematic and includes
the components listed in Table~\ref{systmatic-f}. These
are quoted as percentages of the
branching fraction itself.  The third error is propagated from the statistical
error on the $D^0$, $D^+$ and $D^{*+}$ decay branching fractions.

\section{ Polarization}
\label{POLA} 

The polarization of $D^*$ mesons has been predicted for semileptonic $B$
decays and for two body hadronic decays \cite{WWU,ROSN}.  A model dependent
estimate of momentum dependence of the polarization for directly produced
$D^*$ mesons is available for inclusive decays \cite{WWU}.  CLEO has
previously measured the $D^{*+}$ polarization \cite{INEX}.  Here we present a
new measurement of the $D^{*+}$ polarization with higher statistics. However,
the present detector is operated with a higher magnetic field that makes
impossible to extend the measurement to low $D^{*+}$ momenta for the
$D^{*+}\to D^0\pi^+$ decay mode.  We also present the first measurement of
$D^{*0}$ polarization in inclusive $B$ decays.  In this case, the measurement
can be extended to the lowest momenta without difficulty.  These polarization
measurements served also as a check of the accuracy of our Monte Carlo
simulation of $B$ decay.  

The polarization as a function of $x$ is determined from the distribution of
the $D^*$ decay angle, $\theta$.  This is the angle between the direction of
flight of the $D^{*}$ in the laboratory frame and the direction of the
daughter $D^0$ in the $D^*$ rest frame.  The distribution of this decay angle
can be parameterized as
\begin{equation}
{{dW} \over {d\cos\theta}}= {{3\over{6+2\alpha}}[1+\alpha\cos^2\theta]} 
\label{COSTH}\end{equation}

This is equivalent to the expression in terms of the spin-density matrix
element $\rho_{00}$:
\begin{equation}
W(\cos\theta)={3\over{4}}[(1-\rho_{00})+(3\rho_{00}-1)\cos^2\theta]
\end{equation}
Here, the spin-density matrix is determined in a coordinate system with the
quantization axis along the direction of motion of $D^*$ in the laboratory
frame. The element $\rho_{00}$ is the probability for the $D^*$ to be in the
$J_z = 0$ state.

The polarization parameter $\alpha$ is related to the longitudinal and
transverse decay rates of $D^*$ as 
$$\alpha = {\Gamma_L\over\Gamma_T}-1$$ 

When $\alpha$ is close to its lower bound of $-1$, the distribution is
$\sin^2\theta$ implying transverse polarization while large values of
$\alpha$ imply longitudinal polarization.

Only the $D^{*} \rightarrow D^0 \pi, D^0 \rightarrow K^- \pi^+$ decay mode is
used.  We applied the same selection criteria that were used in the branching
fraction measurements to obtain the $D^*$ samples.  The sample was divided
into 5 intervals in $0.0<x<0.5$ and in 5 intervals in $\cos\theta$.  Each of
the $K\pi$ distributions was fitted to a Gaussian shape plus polynomial
background.  We repeat the same analysis
procedure on the Monte Carlo simulated data to find the dependence of
efficiencies on $x$ and $\cos\theta$.

\begin{figure}[htbp]
\centerline{\psfig{file=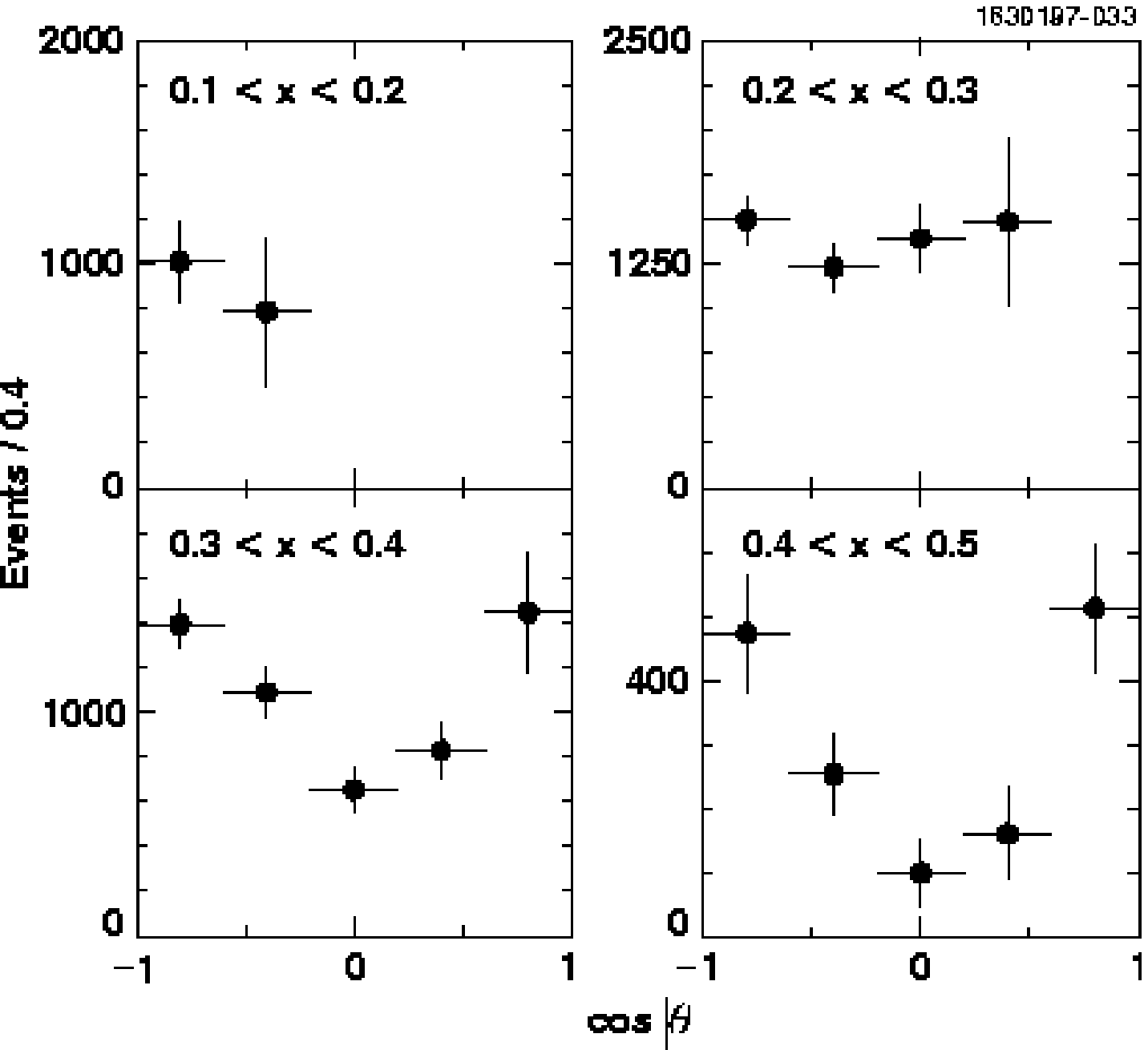,width=3.5in}}
\caption{The background subtracted and efficiency corrected yield of $D^{*+}$
as a function of $\cos \theta$ for four $x$ bins from 0.1 to 0.5.} 
\label{pol1}
\end{figure}

\begin{figure}[htbp]
\centerline{\psfig{file=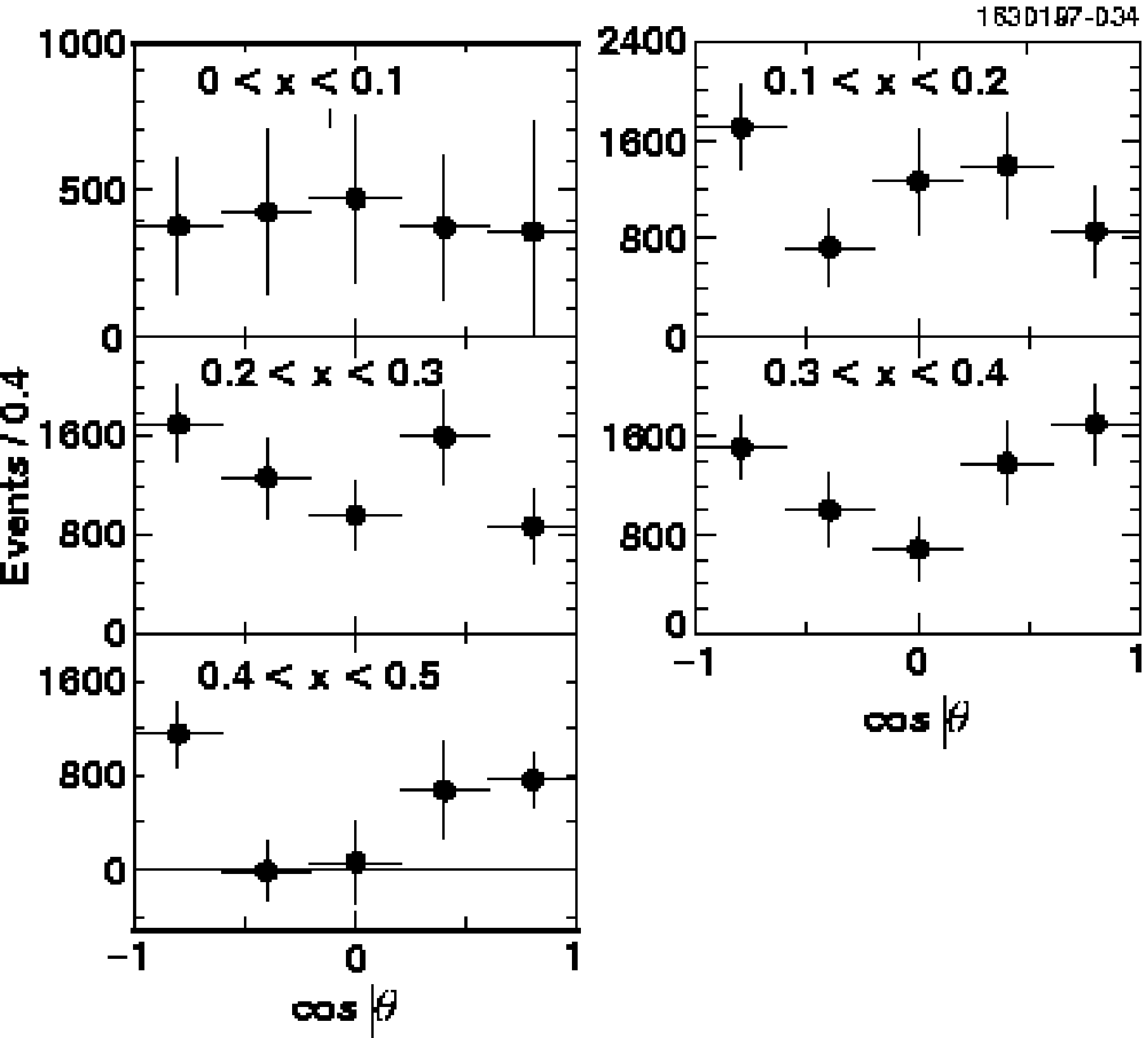,width=3.5in}}
\caption{The background subtracted and efficiency corrected yield of $D^{*0}$
as a function of $\cos \theta$ for five $x$ bins from 0.1 to 0.5.} 
\label{pol2}
\end{figure}

Figures~\ref{pol1} and \ref{pol2} show the efficiency corrected and
background subtracted $\cos\theta$
distributions for $D^{*+}$ and $D^{*0}$ respectively.  We have compared these
distributions with those predicted by our Monte Carlo simulation. The
simulation appears to correctly model the data. This checks our results on
spectra and $B$ decay branching ratios because an incorrect simulation of the
polarization may result in an incorrect determination of the detection
efficiency.

\begin{figure}[htbp]
\centerline{\psfig{file=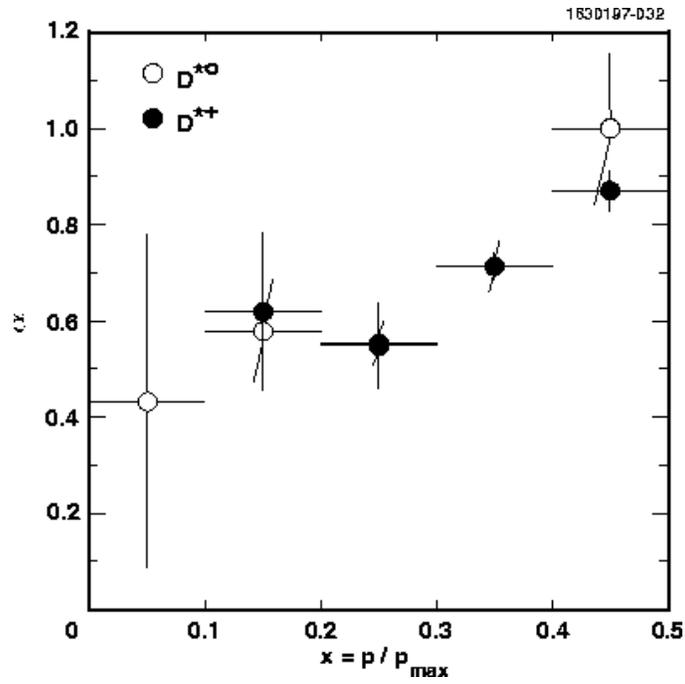,width=3.5in}}
\caption{$\alpha(D^{*})$ vs $x=p/p_{max}$
($p_{max}=4.950~GeV/c$).}\label{pol3} 
\end{figure}

Figure \ref{pol3} shows $\alpha$ as a function of the scaled momentum
variable $x$ for $D^{*+}$ and $D^{*0}$ .

\section{Discussion and Summary}
In Table~\ref{prev.res.} we compare our results with those from previous
measurements.  In order to make the comparison independent of $D^*$ and $D$
decay branching fractions used in the different experiments, we give the
product of the branching fractions.

Only the statistical error is reported here for our current results (third
column) 
because the systematic error on these products of branching ratios
cannot take advantage of some cancellations when we divide them by $D$ and
$D^*$ branching ratios measured in our own experimant.  Our results are
consistent with previous measurements, except for the $B\rightarrow D^0 X$
case where our branching fraction is significantly higher.  The branching
fraction ${\cal B}(B\to D^{*0} X)$ is measured here for the first time.  The
mode ${\cal B}(B\to D^{*+} X)$ is reconstructed in the $D^{*+}\to D^+\pi^0$
decay mode for the first time.  The result for ${\cal B}(B\to D^{*+} X)$ in
the $D^{*+}\to D^0\pi^+$ decay mode refers only to $x_D > 0.15$ and the
previous one from CLEO only to $x_D > 0.10$.  Table~\ref{brtable} summarizes
the results on the inclusive $B$ decay branching fractions.

From the measurement of ${\cal B}(B\to D^{*0} X)$ we
can determine the ratio 
$${{\cal B}(B\to D^{*0} X)\over {\cal B}(B\to D^{*+} X)}= 1.03\pm 0.07 \pm
0.09 \pm 0.08$$ 
which is consistent with the naive expectation of 1.00.

Using these new measurements, and previous ones shown in
Table~\ref{charm-count} , we can now calculate the average number of charm
quarks produced in B decay,

$$ \langle n_c\rangle=1.10\pm 0.05,$$ 

This value is consistent with the naive expectation of 1.15, but it is
considerably lower than the value ($\sim 1.30$) required to account for the
low value of the $B$ semileptonic branching ratio in models where the channel
$b\rightarrow c\overline c s$ is enhanced
\cite{FWD,BAG1,NEUB,HITO,BLOK,SIMU}.


\centerline{\bf ACKNOWLEDGEMENTS}
\smallskip
We gratefully acknowledge the effort of the CESR staff in providing us with
excellent luminosity and running conditions.
J.P.A., J.R.P., and I.P.J.S. thank                                           
the NYI program of the NSF, 
M.S. thanks the PFF program of the NSF,
G.E. thanks the Heisenberg Foundation, 
%
%
K.K.G., M.S., H.N.N., T.S., and H.Y. thank the
OJI program of DOE, 
J.R.P., K.H., M.S. and V.S. thank the A.P. Sloan Foundation,
and A.W. and R.W. thank the 
Alexander von Humboldt Stiftung
for support.
%
M.S. is supported as a Scholar of the Cottrell Research Corporation.
This work was supported by the National Science Foundation, the
U.S. Department of Energy, and the Natural Sciences and Engineering Research 
Council of Canada.



\begin{center}
  TABLES
\end{center}

\begin{center} 
\begin{table}[h] 
\caption{
\label{tab:dm}
$\delta m$ cut and width of the sidebands (in MeV).
}
\begin{tabular}{ccccc}
channel  & width & signal region &  lower-band & upper-band\\
\hline  
$D^{*+} \rightarrow D^{0}\pi^{+}$ & $\pm 1.5$ & 144.1-147.1 & 141.1-142.6 & 148.6-150.1  \\ 
$D^{*+} \rightarrow D^{+}\pi^{0}$ & $\pm 1.5$ & 139.1-142.1 & 136.1-137.6 & 143.6-145.1  \\
$D^{*0} \rightarrow D^{0}\pi^{0}$ & $\pm 2.0$ & 140.6-144.6 & 137.6-139.6 & 145.6-147.6 \\ 
\end {tabular} 
\end{table} 
\end{center}

\begin{table}[htb]
\caption{\label{systmatic-d0} Relative Systematic Errors on ${\cal B}(B\to
D^0 X)$} 
\begin{center} \begin{tabular}{cc}
residual track finding efficiency uncertainty       &  $ 1.0\%$\\
variation of track quality and geometrical cuts     &  $ 0.5\%$\\
choice of signal shape parameterization             &  $ 2.0\%$\\
error in particle identification efficiency         &  $ 0.8\%$\\
error in number of $B$ and $\bar B$                 &  $ 1.8\%$\\
statistical error on efficiency                     &  $ 0.4\%$\\ \cline{2-2}
Total						    &  $ 3.0\%$\\
\end {tabular}   \end{center}  \end{table}

\begin{table}[htb]
\caption{\label{systmatic-dp} Relative Systematic Errors on ${\cal B}(B\to
D^+ X)$} 
\begin{center} \begin{tabular}{cc}
residual track finding efficiency uncertainty       &  $ 1.5\%$\\
variation of track quality and geometrical cuts     &  $ 1.2\%$\\
choice of signal shape parameterization             &  $ 2.0\%$\\
error in number of $B$ and $\bar B$                 &  $ 1.8\%$\\
statistical error on efficiency                     &  $ 0.7\%$\\
uncertainty in background shape                     &  $ 1.7\%$\\
smoothing of the efficiency vs $x$                  &  $ 0.14\%$\\
estimate of kinematical reflections                 &  $ 0.10\%$\\ \cline{2-2}
Total						    &  $ 3.8\%$\\
\end {tabular}   \end{center}  \end{table} 

\begin{table}[htb]
\caption{\label{systmatic-f0} Relative Systematic Errors on ${\cal B}(B\to
D^{*0} X)$} 
\begin{center} \begin{tabular}{cc}
residual particle finding efficiency uncertainty    &  $ 5.1\%$\\
variation of track quality and geometrical cuts     &  $ 0.5\%$\\
choice of signal shape parameterization             &  $ 2.0\%$\\
error in number of $B$ and $\bar B$                 &  $ 1.8\%$\\
statistical error on efficiency                     &  $ 2.1\%$\\
choice of the $\delta m$ region widths              &  $ 1.2\%$\\
scale factor in the $\delta m$ sideband subtraction &  $ 3.3\%$\\  \cline{2-2}
Total						    &  $ 7.1\%$\\
\end {tabular}   \end{center}  \end{table}

\begin{table}[htb]
\caption{\label{systmatic-f} Relative Systematic Errors on ${\cal B}(B\to
D^{*+} X)$} 
\begin{center} \begin{tabular}{cc}
residual track finding efficiency uncertainty 		& $ 5.2\%$\\ 
variation of track quality and geometrical cuts 	& $ 0.5\%$\\ 
choice of signal shape parameterization 		& $ 2.0\%$\\ 
error in number of $B$ and $\bar B$ 			& $ 1.8\%$\\ 
choice of the $\delta m$ region widths 			& $ 1.2\%$\\ 
statistical error on efficiency (combined) 		& $ 1.0\%$\\ \cline{2-2}
Total						    	& $ 6.1\%$\\
\end {tabular}   \end{center}  \end{table}

\begin{table}[htb] 
\caption{\label{prev.res.} Comparison of our results on ``product branching
ratios'' (in \%) with those of previous measurements.  The error in this
work's result is only the statistical one (see text).}       
\begin{center} \begin{tabular}{cc c c c}  
 Channel & Measured Br. Ratio & This work & CLEO \cite{INEX} & ARGUS
\cite{ARGINC} \\ \hline 
$B\to D^0 X$ 
&${\cal B}_B {\cal B}_D$ & $2.51\pm 0.06$ & $2.33\pm 0.12\pm 0.14$ &
                                       $1.94\pm 0.15\pm 0.25$ \\ 
$B\to D^+ X$ 
&$ {\cal B}_B {\cal B}_D$ & $2.16\pm 0.08$ & $2.26\pm 0.30\pm 0.18$ & 
                                       $2.09\pm 0.27\pm 0.40$ \\
$B\to D^{*0} X$ &${\cal B}_B {\cal B}_{D^*} {\cal B}_D$ & $0.620\pm 0.031$ &
-- &  -- \\ 
$B\to D^{*+} X$ &${\cal B}_B {\cal B}_{D^*} {\cal B}_{D^0}$ & $0.570\pm
0.021$ & $0.56\pm 0.03\pm 0.05$ & $0.71\pm 0.06\pm 0.12$ \\ 
$B\to D^{*+} X$ &${\cal B}_B {\cal B}_{D^*} {\cal B}_{D^+}$ & $0.639\pm
0.054$ &                        ---  & --- \\ 
\end {tabular}   \end{center}  \end{table}

\begin{table}[hbt] 
\caption{\label{brtable} 
Inclusive $B$ decay branching fractions to $D$ and $D^*$.}
\begin{center} \begin{tabular}{cc} 
${\cal B}(B\to D^0 X)$    & $0.636 \pm 0.014 \pm 0.019 \pm 0.018$ \\ 
${\cal B}(B\to D^+ X)$    & $0.235 \pm 0.009 \pm 0.009 \pm 0.024$ \\
${\cal B}(B\to D^{*0} X)$ & $0.247 \pm 0.012 \pm 0.018 \pm 0.018$ \\ 
${\cal B}(B\to D^{*+} X)$ & $0.239 \pm 0.015 \pm 0.014 \pm 0.009$ \\ 
\end {tabular}   \end{center}  \end{table}


\begin{table}[hbt] 
\caption{\label{charm-count} Charm counting}  
\begin{center} \begin{tabular}{cc}   
${\cal B}(\bar{B}\to D^0 X)+{\cal B}(\bar{B}\to D^+ X)$ & $0.871\pm 0.035$ \\
${\cal B}(\bar{B}\to D_s X)$
\footnote{We took the product of branching ratios measured by CLEO
\cite{DSMEN}:  ${\cal B}(\bar{B}\to D_s X)\cdot{\cal B}(D_s^+\to \phi\pi^+) =
(0.424\pm 0.014\pm 0.031)\%$ and divided it by the world average ${\cal
B}(D_s^+\to \phi\pi^+) = (0.36\pm 0.09)\%$ \cite{PDG}.}   & $0.118\pm 0.031$ \\
${\cal B}(\bar{B}\to\Lambda_c X)$
\footnote{We have used the product branching ratio ${\cal B}(B\to\Lambda_c\
X){\cal B}(\Lambda_c \to p K^- \pi^+) = (0.181\pm 0.022\pm 0.024)\% $ from
\cite{ZOELLER} and divided it by the world average ${\cal B}(\Lambda_c
\to p K^- \pi^+) = (4.4 \pm 0.6)\%$ \cite{PDG}.}	& $0.039\pm 0.020$ \\
${\cal B}(\bar{B}\to\Xi^+_c X)$
\footnote{This branching fraction and the following one are derived from the
product branching ratios reported in \cite{CLEO95A}.  The ${\cal
B}(B\to\Xi_c\ X)$ branching fractions in these papers were calculated under
the assumptions that $\Gamma(\Xi_c\to X\ell\bar{\nu}) = \Gamma(D\to
X\ell\bar{\nu})$.  M.B. Voloshin \cite{VOLO} estimated that, because of Pauli
interference of the strange quark, $\Gamma(\Xi_c\to X\ell\bar{\nu})$ is in
fact larger.  His estimate (M.B. Voloshin, private communication) is
$\Gamma(\Xi_c\to X\ell\bar{\nu}) = (2.5\pm1.0) \Gamma(D\to X\ell\bar{\nu})$.
Furthermore, here we assume $\Gamma(\Xi_c\to\Xi\ell\bar{\nu})/\Gamma(\Xi_c\to
X\ell\bar{\nu}) = 0.8\pm0.2$ rather that 1.0 as in ref.~\cite{CLEO95A}.
Altogether, the values of ${\cal B}(\Xi_c^+\to \Xi^-\pi^+\pi^+)$ and ${\cal
B}(\Xi_c^0\to \Xi^-\pi^+)$ are decreased by a factor of $2.0\pm0.9$ with
respect to those in ref.~\cite{CLEO95A}.}	& $0.008\pm 0.005$ \\ 
${\cal B}(\bar{B}\to\Xi^0_c X)$			& $0.012\pm 0.009$ \\ 
$2\times{\cal B}(\bar{B}\to {\rm\ Charmonia\ } X)$
\footnote{We took the inclusive, direct B decay branching fractions to
$\psi,\ \psi\prime,\ \chi_{c1}$ and $\chi_{c2}$ from the CLEO result
\cite{BALEST} and added $0.009\pm 0.003$ (the upper limit on ${\cal B}(B\to
\eta_c\ X)$) to take into account all other charmonium production from $B$
decay.} 					& $0.054\pm 0.007$ \\
\end {tabular}   \end{center}  \end{table}

\newpage
\begin{center}
  FIGURES
\end{center}


\begin{thebibliography}{99}

\begin{center}
REFERENCES
\end{center}

\def\etal{{\sl et al.}}
\def\PRref#1&#2&#3(#4){\unskip\ #1~\bf #2\rm, #3 (#4)}
\def\PRL{{\it Phys.\ Rev.\ Lett.}}
\def\PRD{{\it Phys.\ Rev.}\ {\bf D}}
\def\PLB{{\it Phys.\ Lett.}\ {\bf B}}
\def\NPB{{\it Nucl.\ Phys.}\ {\bf B}}
\def\ZPC{{\it Z.\ Phys.}\ {\bf C}}
\def\NIM{{\it Nucl.\ Inst.\ and Meth.}\ {\bf A}}

\bibitem{SLS}
For a review of the experimental and theoretical status of $B$ meson decays,
see``B Decays'', S. Stone ed., 2nd Ed., World Scientific, Singapore,
1994; ``B Mesons'', T.E.~Browder and K.~Honscheid, to appear in {\it Progress
in Nuclear and Particle Physics}, Vol.~35, 81-220 (1995).

\bibitem{FWD}
A.F. Falk, M.B. Wise, I.Dunietz, 
\PRref\PRD&51&1183 (1985).

\bibitem{WWU}
M. Wirbel and Y.-L. Wu\PRref\PLB&228&430 (1989).

\bibitem{SVsemi}
I.I. Bigi, M. Shifman, N.G. Uraltsev and A.I. Vainshtein, {\it Phys.\  
Rev.\ Lett.}\ {\bf 71}, 496 (1993);  B. Blok, L. Koyrakh, M. Shifman and
A.I. Vainshtein, {\it Phys.\ Rev.}\ {\bf D49}, 3356 (1994).

\bibitem{MW}
A. Manohar and M.B. Wise, {\it Phys.\ Rev.}\ {\bf D49}, 1310 (1994).

\bibitem{Mannel}
T. Mannel, {\it Nucl.\ Phys.}\ {\bf B413}, 396 (1994).

\bibitem{tau}
A.F. Falk, Z. Ligeti, M. Neubert and Y. Nir, {\it Phys.\ Lett.}\ {\bf B326},
145 (1994); L. Koyrakh, {\it Phys.\ Rev.}\ {\bf D49}, 3379 (1994);
S. Balk, J.G. K\"orner, D. Pirjol and K. Schilcher, {\it Z.Phys.} {\bf C64},
37 (1994).

\bibitem{SVrare} 

I.I. Bigi, N.G. Uraltsev and A.I. Vainshtein, {\it Phys.\ Lett.}\ {\bf B293},
430 (1992); Erratum, {\it Phys.\ Lett.}\ {\bf B297}, 477 (1993); I.I.~Bigi,
B. Blok, M. Shifman, N.G. Uraltsev and A.I. Vainshtein, 7th Meeting of the
Division of Particles Fields of the APS (DPF 92), Batavia, IL, 10-14 Nov
1992, published in DPF Conf.1992:610-613.

\bibitem{BS}
B. Blok and M. Shifman, {\it Nucl.\ Phys.}\ {\bf B399}, 441 (1993); B. Blok
and M.~Shifman, {\it Nucl.\ Phys.}\ {\bf B399}, 459 (1993); N. Bilic,
B. Guberina and J. Trampetic, {\it Nucl.\ Phys.}\ {\bf B248}, 261 (1984);
M. Voloshin and M. Shifman, {\it Sov.\ J.\ Nucl.\ Phys.}\ {\bf 41}, 120
(1985); M. Voloshin and M. Shifman, {\it Sov.\ Phys.--JETP} {\bf 64}, 698
(1986); V.A. Khoze, M. Shifman, N.G. Uraltsev and M. Voloshin, {\it Sov.\ J.\
Nucl.\ Phys.}\ {\bf 46}, 112 (1987).

\bibitem{BAG1} E.  Bagan \etal, \PRref\NPB&432&3 (1995);
\PRref\PLB&342&362 (1995); \PRref\PLB&351&546 (1995), 
[E: \PRref\PLB&374&363 (1996)]. 

\bibitem{NEUB}
M.  Neubert and C.T.  Sachrajda, ``Spectator Effects in Inclusive
Decays of Beauty Hadrons'', preprint CERN-TH/96-19 (hep-ph/9603202), submitted
to {\it Nucl. Phys.} {\bf B}.

\bibitem{HITO} G. Buchalla, I. Dunietz and H. Yamamoto, {\it Phys. Lett.}
{\bf B364}, 188 (1995).

\bibitem{BLOK} 
B. Blok and T. Mannel, {\it Mod. Phys. Lett.} {\bf A11}, 1263, (1996).

\bibitem{SIMU}
I.L. Grach, I.M. Narodetskii, S. Simula and K.A. Ter-Martirosyan,
``Exclusive and Inclusive Weak Decays of the B-Meson'', Istituto Superiore di
Sanita', preprint no. INFN-ISS 95/17 (hep-ph/9603239) submitted
to {\it Nucl. Phys.} {\bf B}.

\bibitem{INEX}
D.~Bortoletto \etal\ (CLEO), \PRref\PRD&45&21 (1992).

\bibitem{ARGINC}
H.~Albrecht \etal\ (ARGUS), \PRref\ZPC&52&353 (1991).

\bibitem{BPBO}
B. Barish \etal\ (CLEO) \PRref\PRD&51&1014(1995).

\bibitem{DET}
Y. Kubota \etal\ (CLEO), \PRref\NIM&320&66(1992).

\bibitem{DRII}
D.G. Cassel \etal, \PRref\NIM&252&325(1986).

\bibitem{CSI}
E. Blucher \etal, \PRref\NIM&249&201 (1989).

\bibitem{MORR}
R. Morrison \etal\ (CLEO), \PRref\PRL&67&1696 (1991).

\bibitem{MUii}
D. Bortoletto \etal, \PRref\NIM&320&114 (1992).

\bibitem{SJO}
T. Sj\"ostrand, Comp. Phys. Comm. {\bf 39}, 347 (1986), T. 
Sj\"ostrand, M.~Bengston, Comp. Phys. Comm. {\bf 43}, 367 (1987).

\bibitem{DDK} Our Monte Carlo simulation does not include the possibility of
$B\to D \bar{D} K X$ decay.  Since our detection efficiency has been obtained
bin-by-bin in the $D$ momentum, this omission does not affect our result.

\bibitem{GEANT}
R. Brun \etal, GEANT 3.14, CERN DD/EE/84-1.

\bibitem{FOX}
G.C. Fox and S. Wolfram, \PRref\PRL&41&1581(1978).

\bibitem{BUT}
F. Butler \etal\ (CLEO), \PRref\PRL&69&2041(1992).

\bibitem{AKER}
D. Akerib \etal\ (CLEO), \PRref\PRL&71&3070(1993). We use the value 3.91\%
before radiative corrections because our $D^0$ yield is not corrected for
radiative corrections. 

\bibitem{CINAB}
D. Cinabro \etal\ (CLEO), \PRref\PRL&72&1406(1994)

\bibitem{BALE}
R. Balest \etal\ (CLEO), \PRref\PRL&72&2328(1994).

\bibitem{PDG}
Review of Particle Properties, \PRref\PRD&54&1(1996).

\bibitem{BARISH}
For a discussion of charged track and $\pi^0$ detection efficiency and their
errors see ref. \cite{BPBO}

\bibitem{BKT}
CLEO work in preparation.

\bibitem{DSMEN}
D.~Gibaut \etal\ (CLEO), \PRref\PRD&53&4734 (1996).

\bibitem{MWOOD}
CLEO internal note CBX 96-43

\bibitem{ROSN}
J.L. Rosner \PRref\PRD&42&3732(1990).

\bibitem{ZOELLER} M. Zoeller, Ph.D. dissertation, S.U.N.Y. at Albany, (1994)

\bibitem{CLEO95A} D. Cinabro \etal\ (CLEO), paper contributed to the ICHEP
conference, Glasgow, 1994, preprint CLEO CONF 94-08, and J.P.  Alexander
\etal\ (CLEO) \PRref\PRL&74&3113(1995), E: \PRref\PRL&75&4155(1995).

\bibitem{VOLO} M.B. Voloshin, {\it Phys.Lett.} {\bf B385}, 369 (1996).

\bibitem{BALEST} R. Balest \etal\ (CLEO) {\it Phys. Rev.} {\bf D52}, 2661
(1995). 

\end{thebibliography}
\end{document}